\documentclass[twocolumn,aps,prb,groupedaddress,showpacs, superscriptaddress]{revtex4-2} 
\usepackage{amsfonts}
\usepackage{amsmath}
\usepackage{amssymb}
\usepackage{txfonts}
\usepackage{tikz}
\usetikzlibrary{shapes.geometric}
\usepackage{pgffor}
\usepackage{pxfonts}
\usepackage{graphicx,bm,units,yfonts}
\usepackage{subfigure}
\usepackage{hyperref}
\usepackage{verbatim}
\usepackage{listings}
\usepackage[compat=0.4]{yquant}
\usepackage[braket, qm]{qcircuit}
\usepackage{slashbox}
\usepackage{lipsum}

\newcommand{\AM}{{\mathbb A}}
\newcommand{\BM}{{\mathbb B}}
\newcommand{\CM}{{\mathbb C}}
\newcommand{\DM}{{\mathbb D}}

\newcommand{\PM}{{\mathbb P}}
\newcommand{\RM}{{\mathbb R}}

\newcommand{\TM}{{\mathbb T}}
\newcommand{\ZM}{{\mathbb Z}}

\newcommand{\EM}{{\mathbb E}}

\newcommand{\Cc}{{\mathcal C}}
\newcommand{\Dd}{{\mathcal D}}

\newcommand{\Ff}{{\mathcal F}}

\newcommand{\Nn}{{\mathcal N}}

\newcommand{\Ll}{{\mathcal L}}
\newcommand\myatop[2]{{{#1}\atop#2}}

\begin{document}

\title{Vertex Lattice Models Simulated with Quantum Circuits}

\author{Jechiel Van Dijk}

\affiliation{Department of Physics, Yeshiva University, New York, NY 10016, USA}

\author{Emil Prodan}

\affiliation{Department of Physics, Yeshiva University, New York, NY 10016, USA}

\begin{abstract}
Classical planar vertex models afford transfer matrices with real and positive entries, which makes this class of models suitable for quantum simulations. In this work, we support this statement by building explicit quantum circuits that implement the actions of the transfer matrices on arbitrary many-qubit states. The number of qubits and the depth of the circuits grow linearly with the size of the system. Furthermore, we present tests using quantum simulators and demonstrate that important physical quantities can be extracted, such as the eigen-vector corresponding to the largest eigenvalue of the transfer matrix and the ratio of the second to first largest eigenvalue. Challenges steaming from the non-unitarity of the transfer matrix are discussed.
\end{abstract}

\maketitle

\section{Introduction} 

One of the common features of the correlated physical systems, both classical and quantum, is a configuration space whose complexity grows exponentially with the size of the system. Furthermore, by definition, the state of a correlated system is not a simple product state and these specific characteristics make the simulations of these systems extremely demanding.

When fully developed, the quantum computers will supply a better hardware match for such problems \cite{FeynmanIJTP1982}. Of course, this does not imply that any correlated system will be solved with ease on these platforms. However, specific classes of correlated systems are already known to be approachable by quantum computers. One such class is that of systems affording matrix product states \cite{VidalPRL2003,QiskitMPS}, which can be simulated with quantum circuits that grow linearly with the size of the systems. Many other classes have been identified in the past few years \cite{Raeisi2012,McCleanNJP2016,JiangPRA,ChildsPRL2019,Babbush2018,
KivlichanQ2020,Kaicher,RahmaniPRXQ2020,AruteScience2020,
StengerPRR2020,BarratNPJQ2021,Clinton2021} and, definitely, the efforts on simulating correlated systems with quantum algorithms is gaining traction.

In the present work, we identify yet another class of correlated systems that are particularly good candidates for quantum simulations. These are the physical systems that afford a transfer matrix. Our main observation is that the states of these systems can be thought of as a kind of {\it non-commutative} product and products of matrices can be efficiently implemented and simulated with quantum circuits. As an example, we focus here on the classical planar vertex models, whose partition functions, expectation values of the physical observables and correlation functions can be calculated with the aid of a transfer matrix \cite{BaxterBook}. At their turn, the transfer matrices of these particular models are products of the so called $\RM$ matrices and this makes them special because the entire computations reduce to evaluating products of matrices. Such a product contains a number of terms that is proportional with the lateral size of the lattice. As a result, the actions of the transfer matrices can be simulated with circuits whose number of qubits and depth grow linearly with the size of the lattice. However, a challenge still persists for the quantum simulations, spurring from the non-unital character of these $\RM$ matrices.

We will focus here exclusively on the computational aspects, leaving the actual investigation of the vertex models for the future. Specifically, we demonstrate that the action of the transfer matrix on an arbitrary vector can be simulated by circuits that grow linearly with the relevant size of the system. Let us recall that the output of a quantum measurement is a histogram of probabilities and, in order to reproduce the quantum state itself, one needs to employ quantum tomography \cite{DArianoAIEP2003}, which is in general very costly. The transfer matrix of classical vertex models has real and positive entries. Since any quantum state can be decomposed as $\Psi=(\Psi_+ - \Psi_-) + \imath (\Psi'_+ - \Psi'_-)$, where all vectors on the right side have real and positive entries, the action of the transfer matrix can be mapped entirely by acting only on vectors with real and positive entries. If that is the case, then the result is again a vector with real and positive entries. Then the important conclusion is that the action of the transfer matrix can be read off directly from the histograms of the quantum measurements. Hence, classical vertex models are extra-special and the quantum computers could indeed supply an unprecedented  boost to the research of these physical system. For example, the investigation of possible phase transitions requires large system sizes which are prohibitive when approached with classical computers. We should acknowledge, though, that an arbitrary matrix can be always decomposed into four pieces carrying only real or purely imaginary entries with identical signs. Then the action of such matrix on a complex vector can be reconstructed from 16 independent actions of real positive matrices on real positive vectors. In the case studied in this work, these 16 independent actions are reduced to just one action.

As we already mentioned, the $\RM$ matrices are non-unitary. Inspired by the works \cite{TerashimaIJQI2005,QIP162017}, we present a quantum circuit implementation that uses one global ancilla qubit and one projective measurement per $\RM$ matrix. As such, a transfer matrix generated by an $N$ product of $\RM$ matrices can be simulated using just one extra qubit and $N$ projective measurements. The many-qubit state of the circuit reproduces the action of the transfer matrix if and only if all projective measurement return 0. This inherently leads to a dilution of the number of shots, hence, to generate accurate histograms, our protocol requires a number of shots that increases with $N$. As such, our investigation brings out an aspect of quantum computation that received little attention so far, namely, how to ensure, at the hardware level, that the number of shots can be efficiently and reliably increased.

The quantum circuits proposed here supply the actions of the transfer matrices, but only up to a multiplicative factor (see section~\ref{Sec:QC}). This complication is inherent and spurs from the non-unitary character of the $\RM$ matrices. Because of this fact, the circuit does not give us access to the largest eigenvalue of the transfer matrix, which determines the partition function of the system (see sub-section~\ref{Sub-sec:TransfMat}). Nevertheless, the quantum circuit gives us access to the eigen-vector corresponding to the largest eigenvalue, which plays a central role when computing expected values of physical observables (see sub-section~\ref{SubSec:ExpV}). Furthermore, we will show that the proposed quantum circuit also gives us access to the ratio between the second and first largest eigenvalues, which is essential for understanding the asymptotic behavior of the correlation functions (see sub-section~\ref{Sub-sec:CorrF}). 

Performance tests as well as actual results generated with the Qasm quantum simulator are supplied in section~\ref{Sec:Results}. The core Qiskit scripts used in this work are supplied in sections~\ref{Sec:Appendix1} and \ref{Sec:Appendix2} and they are elaborated in section~\ref{Sec:Qiskit}. As we shall see, the codes have a hybrid classical and quantum structure, where the difficult part of evaluating the transfer matrix on a state is sent to the quantum simulator and the returned data is processed classically and further fed to the quantum simulator.

\section{Physical Model and its Statistical Physics}

This section supplies a minimal background on classical planar vertex models and introduces the key aspects that are of interest for a statistical physicist. 

\begin{figure}
\center
  \includegraphics[width=0.99\linewidth]{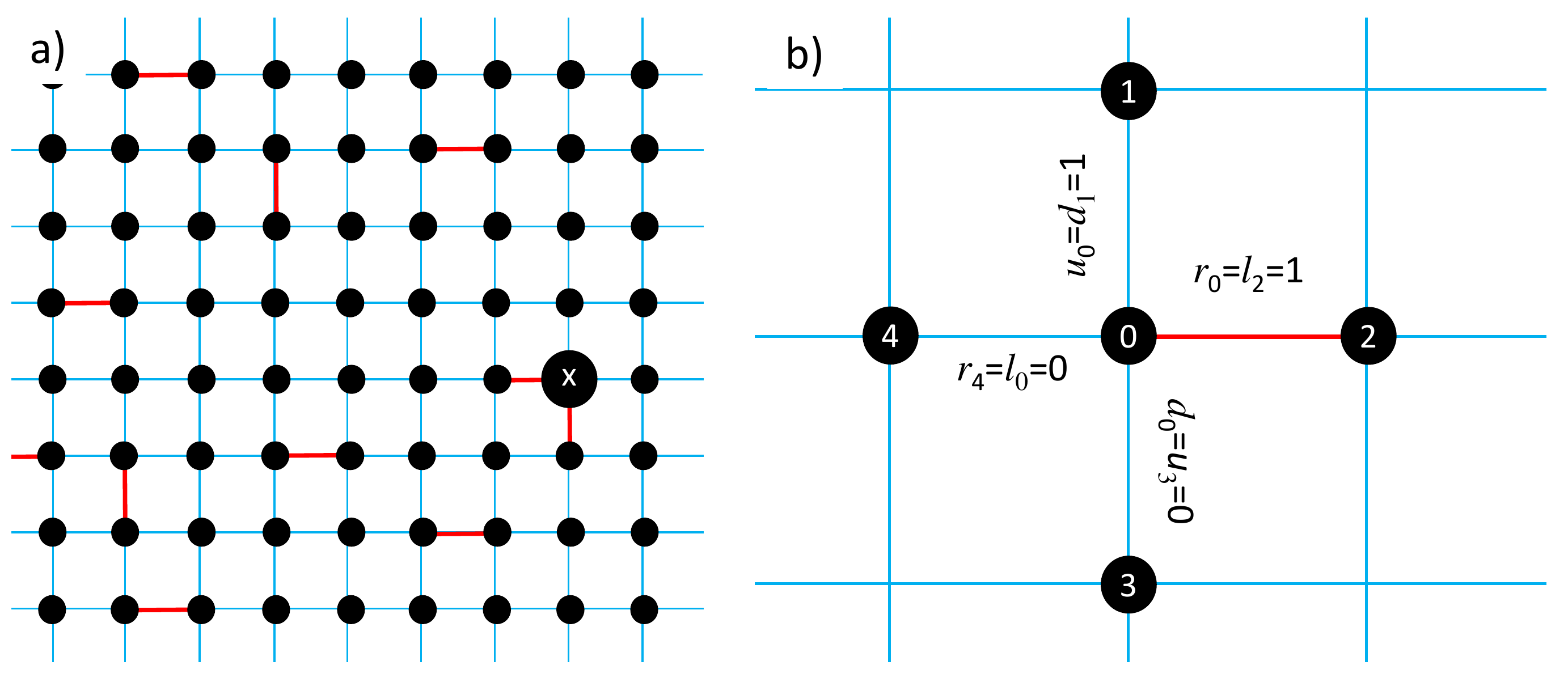}
  \caption{Left: Sample of a square lattice with particles placed at the vertices and connected by two types of bonds, shown in blue and red. The totality of the bonds generates a bond configuration $Q$, which can be thought as a particular coloring of the edges of the lattice. Right: The vertex $0$ surrounded by its four nearest neighbors and with the bonds in a particular configuration. The diagram exemplifies how the indices $d$, $u$, $l$, $r$ work for this particular case.}.
 \label{Fig:PhysicalModel}
\end{figure}

\subsection{The physical model defined}

We will be dealing with the generic 2-dimensional vertex model on the square lattice, which is a familiar physical system to the statistical physicists \cite{BaxterBook}. Still, some readership might come from different backgrounds and, for this reason, we felt compelled to dedicate a section to the model itself and to introduce our notation. It can be summarized as it follows and the reader can consult Fig.~\ref{Fig:PhysicalModel} for guidance:
\begin{enumerate}
\item There exists a lattice $\Ll$ of $N$ columns and $M$ rows, perhaps generated by a trapping potential, which will play no role other than fixing the lattice.
\item At each vertex (or node) of the lattice, there is exactly one particle, just sitting (hence, no kinetics). As such, right from beginning, there are $|\Ll|$ (= cardinal of $\Ll$) particles in the physical system.
\item The particles connect to each other, forming chemical bonds. A pair of neighboring particles can form either a strong bond, hence costing higher energy, or a weak bond, hence costing lesser energy.
\item Each particle has a neighbor in the down (d), up (u), left (l) and right (r) directions. The bonding of a particle with its neighboring particles will be specified by four indices $d,u,l,r$, which take values 0 or 1. For example 
$$
d=0, \ u=1, \ l=1, \ r=0
$$
indicate that the bond with lower neighbor is weak, with the upper neighbor is strong, with left neighbor is strong and with the right neighbor is weak.
\item Each particle contributes to the total energy of the physical system with an amount that is determined by its bondings with the neighboring particles. For the $n$-th particle, this amount is $\epsilon_{d_n}^{u_n}(l_n,r_n)$, where the indices $(u_n,d_n,l_n,r_n)$ communicate how particle $n$ is bonded with its nearest neighbors. 
\item In total, there are $2^4$ possible bonding configurations for each particle, hence the input of the model consists of $2^4$ numerical values:
$$
\epsilon_0^0(0,0), \ \epsilon_1^0(0,0), \ \ldots, \  \epsilon_1^1(1,1).
$$
\item In a particular configuration, the indices $u_n$, $d_n$, $l_n$, $r_n$ can vary from one particle to another, hence they depend on $n$. However, if particle $m$ happens to be to the left of particle $n$, consistency requires that $l_n = r_m$. This also applies to the right, up and down neighbors (see Fig.~\ref{Fig:PhysicalModel}b).
\end{enumerate}

A bond configuration $Q$ is an assignment of 0's and 1's to each bond of the lattice. In Fig.~\ref{Fig:PhysicalModel}, this assignment is communicated by a specific coloring of the network of bonds. If desired, one can think of $Q$ as a point of the set 
$$
\{0,1\}^{N_b}=\{0,1\} \times \{0,1\} \times \ldots \times \{0,1\},
$$
where $N_b$ is the total number of bonds in the system. For example, if the bonds are enumerated in a particular order, then $Q=(0,1,0,\ldots) \in \{0,1\}^{N_b}$ will tell us that the first bond is weak, the second bond is strong, etc.. The outstanding challenge of the problem is that $Q$ can have an awfully large number of different configurations, $2^{N_b}$ to be more precise. For orientation, we indicate that, for a $5 \times 5$ lattice, the number of possible bond configurations is $2^{5\cdot 5} \approx 3.35 \times 10^7$, while for a $10\times 10$ lattice it is $2^{10\cdot 10} \approx 1.26 \times 10^{30}$.

Let us also mention the simple but important fact that a bond configuration $Q$ fixes all the values of the $(d_n,u_n,l_n,r_n)$ coefficients. For example, if the particle $n$ happens to be the particle marked as {\bf x} in Fig.~\ref{Fig:PhysicalModel}a, then
$$
d_n=1 ,\ u_n = 0, l_n = 1, \ r_n = 0. 
$$

We end this sub-section by reminding that the vertex models find applications in areas such as condensed matter physics \cite{Lieb1972}, biophysics \cite{FletcherBJ2014,SilvanusPTRS2017} and chemistry \cite{PaulingJACS1935}.

\subsection{Statistical mechanics considerations}

At a finite temperature, the configuration of the bonds fluctuates in time. If one takes a snapshot of the physical system at time $t$, one could observe a pair of nearest neighboring particles forming a strong bond. However, in a snapshot taken at $t + \Delta t$, one may observe a weak bond between the same pair of particles. For a visual picture, one can imagine Fig.~\ref{Fig:PhysicalModel}a as a dynamical one, where the colors of the bonds change with time. 

If one observes the system over a long enough period of time, one can, at least in principle, build the histogram quantifying the occurrence of each configuration within the time of observation. Statistical mechanics gives us the means to predict how this histogram will look like. Specifically, the probability for a particular configuration $Q$ to occur is given by the Boltzmann weight
\begin{equation}
\PM_\beta(Q) = Z_\beta^{-1} \, e^{-\beta E(Q)}, \quad \beta = \frac{1}{kT},
\end{equation}
where $k$ is Boltzmann's constant, $T$ is the temperature and $E(Q)$ is the total energy of the system for configuration $Q$,
\begin{equation}\label{Eq:QEnergy}
E(Q) = \sum_{n} \epsilon_{d_n}^{u_n}(l_n,r_n),
\end{equation}
with the sum running over all particles in the system. Recall that the values of $(d,u,l,r)$ coefficients are determined by $Q$, for any particle in the system. The constant $Z_\beta$ assures the proper normalization of the probabilities,
\begin{equation}\label{Eq:Partition}
Z_\beta = \sum_Q e^{-\beta E(Q)}.
\end{equation}

The quantity defined in Eq.~\eqref{Eq:Partition} is the partition function, which is central to the statistical physics of the system. It is directly related to the thermodynamic potential $F=U - TS$ called Helmholtz free energy, where $U$ is the internal energy and $S$ is the entropy of the physical system:
\begin{equation}
F = -\beta^{-1} \, \ln \, Z_\beta.
\end{equation}
One of the important tasks of the statistical analysis is to compute the partition function $Z(\beta)$ for a given input of $\epsilon_d^u(l,r)$ of bonding energies.

Another task is mapping the expected values of physical observables. In the present context, the physical observables are simply maps $f(Q)$ from the space of bond configurations to the complex plain. Hence the task is computing
\begin{equation}
\EM(f) : = \sum_Q f(Q) \PM(Q) = Z_\beta^{-1} \sum_Q f(Q) \, e^{-\beta E(Q)}.
\end{equation}
Furthermore, if $\{f_x\}$ is a family of physical observables such that $f_x$ is determined by the configuration of the bonds in a small vicinity of $x \in \Ll$, then one is interested in the correlation function of these observables, namely,
\begin{equation}
C(x,y) = \EM(f_x \cdot f_y) = Z_\beta^{-1} \sum_Q f_x(Q)f_y(Q) \, e^{-\beta E(Q)}.
\end{equation}
As we shall see in section~\ref{Sec:PartFunc}, these quantities of interest can be calculated with the aid of a transfer function.

\subsection{Setting the calculation of the partition function}

As we already mentioned, the sum in Eq.~\eqref{Eq:Partition} involves an exponentially large number of terms, $2^{N_b}$ to be more precise. In this sub-section, we explain how to expand this sum in a manner that will naturally lead us to the concept of transfer matrix. 

Henceforth, let us first take a closer look at the term $e^{-\beta E(Q)}$ and for this we need to be more precise with the labeling of the vertices. Given the 2-dimensional nature of the lattice, we label each vertex by its specific row and column. Henceforth, a vertex labeled by $\bm v = (m,n)$ sits in the $m$-th row, counted from the bottom, and in the $n$-th column, counted from the left. Now, with $E(Q)$ from Eq.~\ref{Eq:QEnergy}, this term takes the form
\begin{equation}
e^{-\beta E(Q)} = e^{-\beta \sum_{\bm v} \epsilon_{d_{\bm v}}^{u_{\bm v}}(l_{\bm v},r_{\bm v})} = \prod_{\bm v} e^{-\beta \, \epsilon_{d_{\bm v}}^{u_{\bm v}}(l_{\bm v},r_{\bm v})}.
\end{equation}
It is convenient to introduce the new quantities
\begin{equation}\label{Eq:RCoeff}
R_{d_{\bm v}}^{u_{\bm v}}(l_{\bm v},r_{\bm v}) : = e^{-\beta \, \epsilon_{d_{\bm v}}^{u_{\bm v}}(l_{\bm v},r_{\bm v})}.
\end{equation}
Then
\begin{equation}\label{Eq:BigProd}
e^{-\beta E(Q)} = \prod_{\bm v} R_{d_{\bm v}}^{u_{\bm v}}(l_{\bm v},r_{\bm v})
\end{equation}
and this is a very complex product. It is advisable to organize the product by rows and columns:
\begin{equation}
\prod_{\bm v} R_{d_{\bm v}}^{u_{\bm v}}(l_{\bm v},r_{\bm v})=\prod_{m=1}^M \prod_{v \in {\rm row}\, m} R_{d_{\bm v}}^{u_{\bm v}}(l_{\bm v},r_{\bm v}).
\end{equation}
We now consider the sum from Eq.~\eqref{Eq:Partition} over the bond configurations $Q$, and notice that it can also be organized in a sum over the vertical bonds and a sum over the horizontal bonds. Then
\begin{equation}\label{Eq:Expansion}
\sum_Q e^{-\beta E(Q)}=\sum_{d_v,u_v}\prod_{m=1}^M  \Big (\sum_{l_v,r_v}^{v \in {\rm row}\,m}\prod_{v \in {\rm row}\, m} R_{d_{\bm v}}^{u_{\bm v}}(l_{\bm v},r_{\bm v}) \Big ).
\end{equation}

\begin{figure}
\center
  \includegraphics[width=0.9\linewidth]{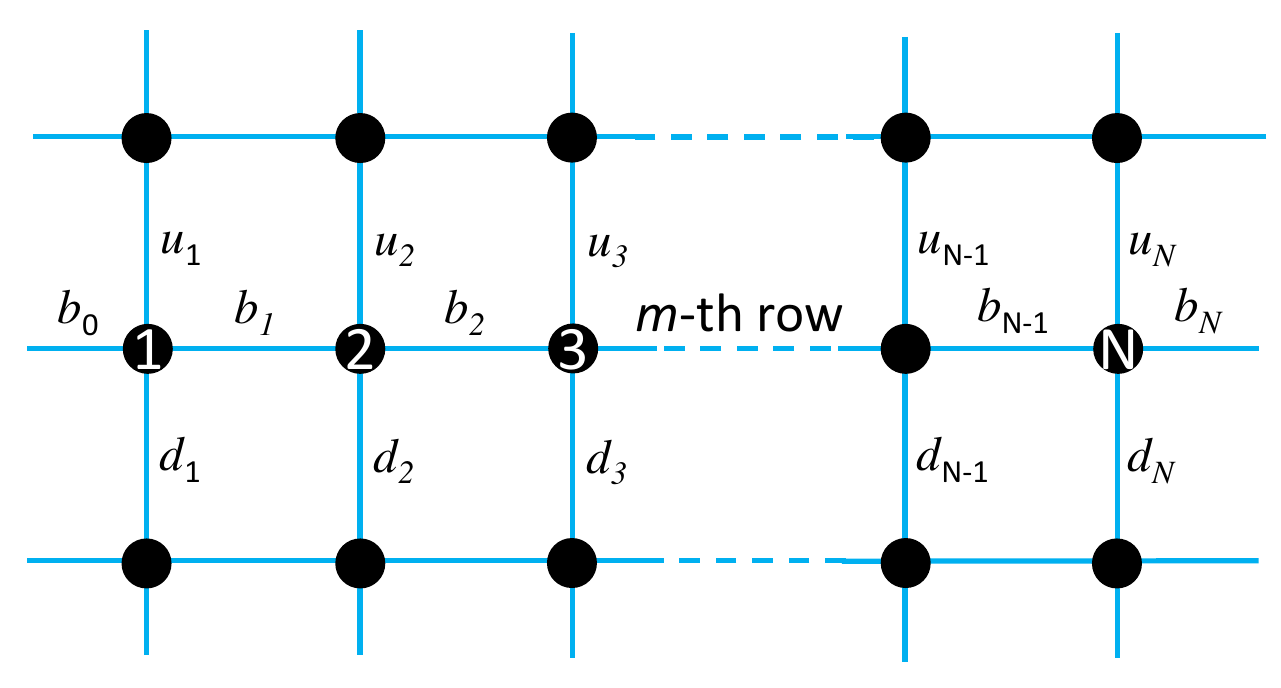}
  \caption{A zoom-in into the $m$-th row of the physical system. The diagram supplies the meaning of the indices used in the calculations carried in the main text, specifically, in Eq.~\eqref{Eq:RowProd}.}
 \label{Fig:RowProd}
\end{figure}

 In Fig.~\ref{Fig:RowProd}, we examine more closely the configurations of a single row and introduce more efficient notation. As one can see, since the row index $m$ is fixed, we erased it entirely from the notation. It will be introduced back when the product over the rows is analyzed. Furthermore, since the bonds need to obey the constraints $r_{m,k}=l_{m,k+1}$, we introduced the symbols $b_k$'s that carry the common values of such pairs of indices. With this notation, 
\begin{equation}\label{Eq:RowProd}
\begin{aligned}
& \qquad \qquad  \sum_{l_v,r_v}^{v \in {\rm row}\,m}\prod_{v \in {\rm row}\, m} R_{d_{\bm v}}^{u_{\bm v}}(l_{\bm v},r_{\bm v}) = \\
& \sum_{b's} R^{u_1}_{d_1}(b_0,b_1) \cdot R^{u_{2}}_{d_{2}}(b_{1},b_{2}) \cdots R^{u_{N}}_{d_{N}}(b_{N-1},b_{N}).
\end{aligned}
\end{equation}

The above expression certainly looks like the product of $N$ matrices. In the following section, we explain how such a chain of products can be computed with the tensor calculus. We deffer the discussion of the product over the row degrees of freedom to the next section. 

\section{Tensor calculus for vertex models}
\label{Sec:TensorCalc}

In the first two sub-sections, we introduce the system of matrix units and demonstrate its effectiveness when it comes to the tensor analysis. This has been already noticed in one of the authors previous work \cite{LiuIJMP2020}.

\subsection{Matrix algebra using the matrix units}

The matrix units for the space of $K \times K$ matrices consist of the elementary matrices $E_{i}^j$, $i,j=\overline{1,K}$, such that $E_{i}^j$ has zero entries except at position $ij$, where the entry is 1. Another way to introduce the matrix units is by the relations
\begin{equation}
V_n^T \cdot E_{i}^j \cdot V_m = \delta_{ni} \delta_{jm},
\end{equation}
where $V_a$ is the column matrix with $K$ entries, of which only the entry at position $a$ is non-zero and equal to 1. The index $a$ takes integer values from $1$ to $K$. Hence, $\{V_a\}$ is the standard basis of $\CM^K$. Since we will deal mostly with $2\times 2$ matrices, we write out the system of units for this case:
$$
E_{1}^1 = {\small \begin{bmatrix}
1&0\\0&0
\end{bmatrix}}, \ 
E_{1}^2 = {\small \begin{bmatrix}
0&1\\0&0
\end{bmatrix}}, \ 
E_{2}^1 = {\small \begin{bmatrix}
0&0\\1&0
\end{bmatrix}}, \ 
E_{2}^2 = {\small \begin{bmatrix}
0&0\\0&1
\end{bmatrix}}.
$$

Obviously, any matrix $A = [a_{ij}]_{i,j=\overline{1,K}}$ can be written as the linear combination
\begin{equation}\label{Eq:Deco1}
A
= a_{11}E_{1}^1 + a_{12}E_{1}^2 + \cdots +  a_{KK}E_{K}^K,
\end{equation}
which, among other things, assures us that the system of unit matrices $\{E_{i}^j\}$ is a basis for the space of $K \times K$ matrices. Throughout, we will adopt Einstein's summation convention, which says that repeating indices are summed over all their allowed values. For example, Eq.~\eqref{Eq:Deco1} simplifies to 
\begin{equation}
A = a_{ij}E_{i}^j.
\end{equation}

From their very definition, one finds the following rule for the multiplication of two matrix units:
\begin{equation}\label{Eq:ERule}
    E_{k}^l\, E_{m}^n =     \begin{cases}
            E_{k}^n, &         \text{if } l=m,\\
            0, &         \text{if } l\neq m.
    \end{cases}
\end{equation}
As a quick application, let us compute
\begin{equation}
A \, B = (a_{ij}E_{i}^j) (b_{nm}E_{n}^m) = a_{ij}b_{nm}E_{i}^jE_{n}^m = a_{ij}b_{jm}E_{i}^m,
\end{equation}
which reproduces the standard multiplication rule of two matrices, $AB =C$, $c_{im}=a_{ij}b_{jm}$. 

\subsection{Tensor products of matrices}

We specialize the discussion to the algebra $M_2(\CM)$ of $2 \times 2$ matrices and introduce the standard qubit basis
\begin{equation}
|0\rangle = {\small \begin{bmatrix} 1 \\ 0 \end{bmatrix}}, \quad |1\rangle = {\small \begin{bmatrix} 0 \\ 1 \end{bmatrix}}.
\end{equation}
Furthermore, we will use the shorthand
\begin{equation}
|i_1 i_2 \ldots i_N \rangle = |i_1\rangle \otimes |i_2\rangle \otimes \cdots \otimes |i_N\rangle, \quad i_k \in \{0,1\},
\end{equation}
for the $N$-th tensor product of vectors.

The tensor products of the unit matrices supply the elementary projection of the tensor space. Indeed,
\begin{equation}\label{Eq:Id1}
\begin{aligned}
 \langle i_1 \ldots i_N | E_{j_1}^{ k_1} \otimes \cdots \otimes E_{j_N}^{ k_N}&|l_1 \ldots l_N \rangle \\
& = \delta_{i_1 j_1} \delta_{k_1 l_1} \cdots \delta_{i_N j_N} \delta_{k_N l_N}
\end{aligned}
\end{equation}
and, since this identity holds for all available values of the indices, it demonstrates that
\begin{equation}\label{Eq:ProjIdentity}
E_{j_1}^{ k_1} \otimes \cdots \otimes E_{j_N}^{ k_N} = |j_1 \ldots j_N \rangle \langle k_1 \ldots k_N|.
\end{equation}

Any element from $M_2(\CM)^{\otimes N}$, that is, any  linear combination of elementary products of matrices
\begin{equation}
\mathbb A = \sum_j A_1^{(j)} \otimes A_2^{(j)} \otimes \cdots A_N^{(j)}, \quad A_n^{(j)} \in M_2(\CM),
\end{equation}
can be written in terms of the matrix units:
\begin{equation}
\mathbb A = A_{i_1 \cdots i_N}^{j_1 \cdots j_N} \, E_{i_1}^{ j_1} \otimes \cdots \otimes E_{i_N}^{j_N},
\end{equation}
where $A_{i_1 \cdots i_N}^{j_1 \cdots j_N}$ are numerical factors. We recall that summation over repeating indices is assumed. The rules addition and multiplication of tensor products become
\begin{equation}
\AM + \BM = (A_{i_1 \cdots i_N}^{j_1 \cdots j_N} + B_{i_1 \cdots i_N}^{j_1 \cdots j_N}) \, E_{i_1}^{ j_1} \otimes \cdots \otimes E_{i_N}^{j_N}
\end{equation}
and
\begin{equation}
\AM \cdot \BM = (A_{i_1 \cdots i_N}^{j_1 \cdots j_N} \, B_{j_1 \cdots j_N}^{k_1 \cdots k_N}) \, E_{i_1}^{k_1} \otimes \cdots \otimes E_{i_N}^{k_N}.
\end{equation}

\subsection{Specialized tensor analysis}
\label{Sec:SpecTens}

We are now ready to describe the computation of Eq.~\eqref{Eq:RowProd} using the tensor calculus. We concentrate on the left side of Eq.~\eqref{Eq:RowProd} and encode all bond configurations of the atom sitting in the $k$-th column in the following tensor product:
\begin{equation}
\mathbb R_{0k} = R_{d_k}^{u_k}(l_k,r_k) \, E_{l_k}^{r_k} \otimes I  \cdots \otimes I \otimes E_{d_k}^{u_k} \otimes I \cdots \otimes I.
\end{equation}
Above, there are exactly $N+1$ matrices in the product, the $E$'s are the matrix units for $2\times 2$ matrices, $E_{l_k}^{r_k}$ sits at position $0$ and $E_{d_k}^{u_k}$ sits at position $k$ in the tensor product and summation over the repeating indices is assumed. It is instructive to compute first the following product:
\begin{equation}
\begin{aligned}
\mathbb R_{01} \mathbb R_{02}  = & R_{d_1}^{u_1}(l_1,r_1) R_{d_2}^{u_2}(l_2,r_2) \,  \\
 & \qquad (E_{l_1}^{ r_1} \otimes E_{d_1}^{ u_1} \otimes I \cdots \otimes I) \\
 & \qquad \qquad (E_{l_2}^{ r_2} \otimes I  \otimes E_{d_2}^{ u_2} \otimes I \cdots \otimes I) \\
 = & R_{d_1}^{u_1}(l_1,r_1) R_{d_2}^{u_2}(l_2,r_2) \\
 & \qquad (E_{l_1}^{ r_1} E_{l_2}^{ r_2} \otimes E_{d_1}^{ u_1}  \otimes E_{d_2}^{u_2} \otimes I \cdots \otimes I)
 \end{aligned}
\end{equation}
Using the rule stated in Eq.~\eqref{Eq:ERule}, we must set $r_1=l_2$ and we denote by $b_1$ the common value. Then
\begin{equation}
\begin{aligned}
\mathbb R_{01} \mathbb R_{02}  =  & R_{d_1}^{u_1}(l_1,b_1) R_{d_2}^{u_2}(b_1,r_2) \\
 & \qquad (E_{l_1}^{ r_2} \otimes E_{d_1}^{ u_1}  \otimes E_{d_2}^{ u_2} \otimes I \cdots \otimes I).
 \end{aligned}
\end{equation}
Throughout, summation over repeating indices is assumed. Then, by iteration,
\begin{equation}\label{Eq:RowProd1}
\begin{aligned}
& \TM:=\mathbb R_{01} \mathbb R_{02} \cdots \mathbb R_{0N}  \\
 & \qquad =  R_{d_1}^{u_1}(l_1,b_1) R_{d_2}^{u_2}(b_1,b_2) \cdots  R_{d_N}^{u_N}(b_{N-1},r_N)\\
 & \qquad \qquad (E_{l_1}^{ r_N} \otimes E_{d_1}^{ u_1}  \otimes E_{d_2}^{ u_2}  \cdots \otimes E_{d_N}^{u_N}).
 \end{aligned}
\end{equation}
The result is a $2^{N+1} \times 2^{N+1}$ matrix $\TM$, written in terms of the unit matrices and having numerical coefficients that reproduce the row products~\eqref{Eq:RowProd} we want to compute.

We proceed now with a computation of the full expansion in Eq.~\eqref{Eq:Expansion}. In the numerical coefficients of Eq.~\eqref{Eq:RowProd1},
\begin{equation}
R_{d_1}^{u_1}(l_1,b_1) R_{d_2}^{u_2}(b_1,b_2) \cdots  R_{d_N}^{u_N}(b_{N-1},r_N),
\end{equation}
the indices $l_1$, $r_N$, $d_1$, \ldots, $d_N$ and $u_1$, \ldots, $u_N$ are un-paired, hence these coefficients are of the form
\begin{equation}\label{Eq:T1}
R_{d_1}^{u_1}(l_1,b_1) \cdots  R_{d_N}^{u_N}(b_{N-1},r_N)={}^1T_{d_1\ldots d_N}^{u_1\ldots u_N}(l_1,r_N),
\end{equation}
a notation we adopt in the following. Hence
\begin{equation}\label{Eq:T1}
\begin{aligned}
 \TM   = &  {}^1T_{d_1\ldots d_N}^{u_1\ldots u_N}(l_1,r_N) \, (E_{l_1}^{ r_N} \otimes E_{d_1}^{ u_1}   \cdots \otimes E_{d_N}^{ u_N}) 
 \end{aligned}
\end{equation}
We now compute
\begin{equation}\label{Eq:ColumnProd1}
\begin{aligned}
 \TM^2   = &  {}^1T_{d_1\ldots d_N}^{u_1\ldots u_N}(l_1,r_N) \, {}^1T_{d'_1\ldots d'_N}^{u'_1\ldots u'_N}(l_1',r_N')\\
 & \qquad (E_{l_1}^{ r_N} \otimes E_{d_1}^{ u_1}   \cdots \otimes E_{d_N}^{ u_N}) \\
 & \qquad \qquad (E_{l'_1}^{ r'_N} \otimes E_{d'_1}^{ u'_1}  \cdots \otimes E_{d'_N}^{ u'_N}) \\
 = &  {}^1T_{d_1\ldots d_N}^{u_1\ldots u_N}(l_1,r_N) \,  {}^1T_{d'_1\ldots d'_N}^{u'_1\ldots u'_N}(l_1',r_N')\\
& \qquad  (E_{l_1}^{ r_N}E_{l'_1}^{ r'_N} \otimes E_{d_1}^{ u_1} E_{d'_1}^{ u'_1}  \cdots \otimes E_{d_N}^{ u_N} E_{d'_N}^{ u'_N})  
 \end{aligned}
\end{equation}
Using again the rules for matrix units multiplications, we see that the following constraints take place,
\begin{equation}
l'_1 = r_N, \ u_1 = d'_1, \ldots, u_N=d'_N.
\end{equation}
The calculation becomes very suggestive if we adopt the following notation for the common value of these indices:
\begin{equation}
s_1 = l'_1 = r_N , \  b_1^1=u_1 = d'_1, \ldots, b_N^1=u_1 = d'_1,
\end{equation}
and use more suggestive symbols $l_1^1 = l_1$ and $r_N^2 = r'_N$. Then
\begin{equation}\label{Eq:ColumnProd1}
\begin{aligned}
 \TM^2  = &  {}^1T_{d_1\ldots d_N}^{b_1^1\ldots b_N^1}(l_1^1,s_1) \,  {}^1T_{b_1^1\ldots b_N^1}^{u'_1\ldots u'_N}(s_1,r_N^2)\\
& \qquad \qquad \qquad \qquad \quad (E_{l_1^1}^{  r_N^2} \otimes E_{d_1}^{ u'_1}  \cdots \otimes E_{d_N}^{  u'_N}). 
 \end{aligned}
\end{equation}
We can also change the notation from $u_i'$ to $u_i$. Then
 \begin{equation}\label{Eq:ColumnProd1}
\begin{aligned}
 \TM^2  = & \, {}^1T_{d_1\ldots d_N}^{b_1^1\ldots b_N^1}(l_1^1,s_1) \,  {}^1T_{b_1^1\ldots b_N^1}^{u_1\ldots u_N}(s_1,r_N^2)\\
& \qquad  \qquad \qquad \qquad \quad (E_{l_1^1}^{  r_N^2} \otimes E_{d_1}^{ u_1}  \cdots \otimes E_{d_N }^{ u_N}) \\ 
 \end{aligned}
\end{equation}
The conclusion is that $\TM^2$ has the same structure as $\TM$ in Eq.~\eqref{Eq:T1},
\begin{equation}\label{Eq:T2}
\begin{aligned}
 \TM^2   = & \, {}^2T_{d_1\ldots d_N}^{u_1\ldots u_N}(l_1^1,r_N^2) \, (E_{l_1^1}^{ r_N^2} \otimes E_{d_1}^{ u_1}   \cdots \otimes E_{d_N}^{ u_N}) 
 \end{aligned}
 \end{equation}
 with
 \begin{equation}
  {}^2T_{d_1\ldots d_N}^{u_1\ldots u_N}(l_1^1,r_N^2) = {}^1T_{d_1\ldots d_N}^{b_1^1\ldots b_N^1}(l_1^1,s_1) \,  {}^1T_{b_1^1\ldots b_N^1}^{u_1\ldots u_N}(s_1,r_N^2).
\end{equation}
This is significant because the calculations can be easily iterated, with the result
\begin{equation}\label{Eq:TM}
\begin{aligned}
 \TM^M   = &  {}^MT_{d_1\ldots d_N}^{u_1\ldots u_N}(l_1^1,r_N^M) \, (E_{l_1^1}^{ r_N^M} \otimes E_{d_1}^{ u_1}   \cdots \otimes E_{d_N}^{ u_N}) 
 \end{aligned}
 \end{equation}
where
 \begin{equation}\label{Eq:TM}
 \begin{aligned}
 &  {}^MT_{d_1\ldots d_N}^{u_1\ldots u_N}(l_1^1,r_N^M) \\
 & \quad = {}^1 T_{d_1\ldots d_N}^{b_1^1\ldots b_N^1}(l_1^1,s_1) \,  {}^1T_{b_1^1\ldots b_N^1}^{b_1^2\ldots b_N^2}(s_1,s_2) \cdots \\
  & \quad \qquad  {}^1T_{d_1\ldots d_N}^{b_1^{M-1}\ldots b_N^{M-1}}(s_{M-2},s_{M-1}) \,  {}^1T_{b_1^{M-1}\ldots b_N^{M-1}}^{u_1\ldots u_N}(s_{M-1},r_N^M)
  \end{aligned}
\end{equation}
If we recall the explicit expression of ${}^1T$ factors, supplied in Eq.~\eqref{Eq:T1}, we see that the product~\eqref{Eq:TM} almost delivers the partition function of the physical system, as formulated in Eq.~\eqref{Eq:Expansion}. This important conclusion is further analyzed in the following section.

\begin{figure}
\center
  \includegraphics[width=0.9\linewidth]{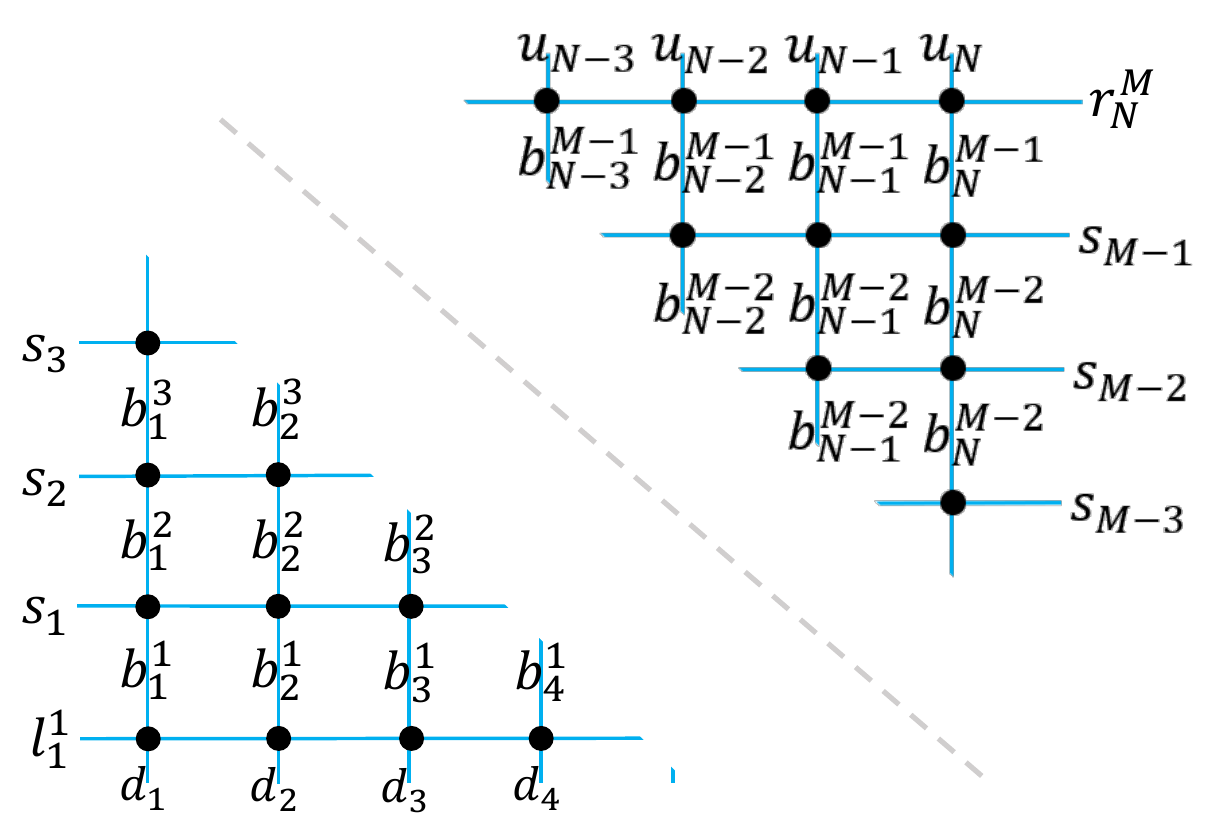}
  \caption{This diagram give a visual representation of the indices appearing in the calculation of the transfer matrix, specifically, in Eq.~\eqref{Eq:TM}.}.
 \label{Fig:FinalProd}
\end{figure}

\section{Transfer matrix}
\label{Sec:PartFunc}

In this section, we supply the connection between the mathematical computation of the previous section and the statistical physics of the vertex model. We also discuss the aspects related to the boundary conditions, convergence rate to the thermodynamic limit and the asymptotic behavior of the correlation functions. The purpose here is to single out various quantities that impact the physics of the vertex lattice and, as such, are interesting to compute.

\subsection{Bulk and boundary degrees of freedom}

To fully understand the expression in Eq.~\eqref{Eq:TM}, we reproduce in Fig.~\ref{Fig:FinalProd} the physical lattice and placed the indices appearing in Eq.~\eqref{Eq:TM} at their rightful place. As one can see, the indices away from the boundary, {\it i.e.} the $b$'s, appear in pairs in Eq.~\eqref{Eq:TM} hence they are all summed up. For the bonds appearing at the lateral boundaries, Eq.~\eqref{Eq:TM} forces the constraints $l_1^{m+1}=r_N^m=s_m$ and $s_m$'s appear in pairs, hence they are also summed up. The only indices that do not appear in pairs are the $d$'s, the $u$'s, $l_1^1$ and $r_N^M$. The conclusion is that Eqs.~\eqref{Eq:TM} and \eqref{Eq:Expansion} are the same except for the contributions of the bonds located at the boundary.

In statistical mechanics, the degrees of freedom are divided into bulk and boundary degrees of freedom. Furthermore, one needs to deal with the inherent physical reality that the boundary degrees of freedom are strongly influenced by the environment. A fundamental principle of thermodynamics is that the ratio $F/|\Ll|$ of the free energy by the particle number converges to a well defined value as $\Ll \rightarrow \ZM^2$, regardless of the conditions imposed on the boundary degrees of freedom. The only exception to this rule happens at the thermodynamic phase transitions. This aspects will be discussed in more details in the following sub-section.

In reality, as well as in our simulations, the physical systems are always finite. Hence, the meaningful quantities to concentrate on are:
\begin{enumerate} 
\item The ratio
\begin{equation}\label{Eq:f}
f(\beta) = F/|\Ll| : = -\beta^{-1} \, \frac{\ln \, Z_\beta}{NM};
\end{equation}
\item Its fluctuations with respect to different boundary conditions one can impose;
\item The rate of convergence to the thermodynamic limit;
\item Expected values of physical observables;
\item Behavior of the correlation functions.
\end{enumerate}

 Given the discussion in the first paragraph, we can make the identification
\begin{equation}\label{Eq:Rel0}
{}^MT_{d_1\ldots d_N}^{u_1\ldots u_N}(l_1^1,r_N^M) = Z_{d_1\ldots d_N}^{u_1\ldots u_N}(l_1^1,r_N^M),
\end{equation}
where on the right we have the partition function of a finite $N \times M$ physical system with the top/bottom boundary degrees of freedom constraint to the $u$ and $d$ values, as well as the corner degrees of freedom constraint at the $l_1^1$ and $r_N^M$ values. Furthermore, a certain type of periodic boundary conditions are imposed on the remaining lateral degrees of freedom. Hence, relation~\eqref{Eq:Rel0} supplies the vehicle to investigate points 1, 2, and 3, above, with respect to boundary conditions on the first and last rows. To investigate the effect of the boundary conditions in the lateral sides of the sample, one can simply rotate the lattice by $90^\circ$ and repeat the analysis.

\subsection{The transfer matrix $\TM$}
\label{Sub-sec:TransfMat}

Given the identity in Eq.~\ref{Eq:Id1}, one can easily establish the following identity:
\begin{equation}\label{Eq:MatElement}
\langle d_N \ldots d_1 d_0|\TM^M |u_1 \ldots u_N u_0 \rangle = Z_{d_1\ldots d_N}^{u_1\ldots u_N}(d_0,u_0).
\end{equation}
This assures us that the partition functions of the physical system with various boundary conditions can all be computed from transfer matrix $\TM$. As already emphasized in our introduction and further discussed below, the transfer matrix $\TM$ contains much more information and, as such, our focus shifts to this object.

Let us acknowledge first that $\TM$ depends entirely on the physical input $\epsilon_d^u(l,r)$ and is not affected by the boundary conditions, as it can be directly seen from its very definition~\eqref{Eq:RowProd1}. The boundary conditions come into play through the many-qubit states in Eq.~\eqref{Eq:MatElement}. With this simple observation, one can give a simple explanation of why the thermodynamic quantities are insensitive to the boundary conditions $\{d_i\}$ and $\{u_i\}$. For this, we will use the spectral decomposition
\begin{equation}\label{Eq:SpecDeco}
\TM = \sum_{j = 0}^{2^{N+1}-1} \Lambda_j \, |\Psi_j^R\rangle \langle \Psi_j^L |,
\end{equation}
where $\Psi_j^{R,L}$ are the left and right eigen-vectors of $\TM$, respectively, which are different from each other for a non-hermitean matrix. Also, the eigenvalues, which are not necessarily real, have been ordered in Eq.~\ref{Eq:SpecDeco} such that $\Lambda_0$ has maximum absolute value. In this specific case, $\Lambda_0$ is always a real quantity. At this point, of course, we assume that $\TM$ is diagonalizable and, since non-diagonalizable matrices form a set of measure zero in the space of matrices, this assumption is not severe at all. We also recall that the left and right eigen-vectors in Eq.~\eqref{Eq:SpecDeco} are normalized such that
\begin{equation}
\langle \Psi_i^L | \Psi_j^R \rangle = \delta_{ij}.
\end{equation}
Now, by taking powers and factoring out the largest eigenvalue,
\begin{equation}
\TM^M = \Lambda_0^M \sum_{j = 0}^{2^{N+1}-1} (\Lambda_j/\Lambda_0)^M \, |\Psi_j^R\rangle \langle \Psi_j^L |,
\end{equation}
one obtains the following asymptotic behavior
\begin{equation}\label{Eq:Asy}
\TM^M = \Lambda_0^M \Big ( |\Psi_0^R\rangle \langle \Psi_0^L | + (\Lambda_1/\Lambda_0)^M \, |\Psi_1^R\rangle \langle \Psi_1^L |+ \ldots \big ).
\end{equation}
If we denote by $P_j = |\Psi_j^R\rangle \langle \Psi_j^L |$ the spectral projection of $\TM$ onto the $j$-th eigenvalue, then all the above can be expressed as
\begin{equation}\label{Eq:Limit1}
\TM^M \approx \Lambda_0^M \, P_0 \quad \mbox{as} \quad M \to \infty.
\end{equation}
Of course, this is valid in general for any matrix. Nevertheless, the conclusion at this point is that
\begin{equation}\label{Eq:Con0}
\lim_{\Ll \to \ZM^2} \frac{\ln \big (\langle \Psi |\TM^M |\Psi' \rangle\big )}{|\Ll|}  = \lim_{N \rightarrow \infty} \frac{\ln (\Lambda_0)}{N},
\end{equation}
for any many-qubit states $\Psi$ and $\Psi'$. As one can see, the contribution of the boundary conditions, which are encoded in $\Psi$ and $\Psi'$, go to zero as
\begin{equation}
\frac{1}{M} \ln \big (\langle \Psi |P_0 |\Psi' \rangle\big ),
\end{equation}
when $M \to \infty$. The conclusion is that
\begin{equation}\label{Eq:Con1}
\lim_{\Ll \to \ZM^2} \frac{\ln \big (Z_{d_1\ldots d_N}^{u_1\ldots u_N}(l_1^1,r_N^M)\big )}{|\Ll|} = \lim_{N \to \infty} \frac{|\Lambda_0|}{N},
\end{equation}
and this explains why the boundary conditions do not have any effect in the thermodynamic limit.

An important and extremely useful piece of information is the rate of convergence of the limit~\eqref{Eq:Con1}. By examining the asymptotic behavior from Eq.~\eqref{Eq:Asy}, we see that this rate is determined by the ratio $\Lambda_1/\Lambda_0$. More precisely, one expects
\begin{equation}\label{Eq:Con2}
\Lambda_0^{-M} \,  \langle \Psi |\TM^M |\Psi' \rangle = \langle \Psi |P_0 |\Psi' \rangle + o\Big ( (\Lambda_1/\Lambda_0)^M \Big ).
\end{equation}
The conclusion is that the ratio $\Lambda_1/\Lambda_0$ dictates the rate of convergence towards the thermodynamic limit with respect to the vertical size of the system. To derive similar statements for the horizontal size of the system, one can simply rotate the system by $90^\circ$, recompute the transfer matrix and map the corresponding ratio $\Lambda_1/\Lambda_0$. 

\subsection{Expected values}
\label{SubSec:ExpV}

We consider here the expected value of a physical observable $f(Q)$, where $f$ is a function which depends on the bond configurations inside a domain $\Dd$ surrounding the central vertex $c$ of the lattice. An example of such observable is $f(Q) = r_{c}$. The expected value of the observable is
\begin{equation}\label{Eq:Exp1}
\begin{aligned}
\EM(f) : & = \sum_{Q} f(Q) \PM_\beta(Q) \\
& = \lim_{\Ll \to \ZM^2} Z_\beta^{-1} \sum_Q f(Q) \, e^{-\beta E(Q)} \\
& =\lim_{\Ll \to \ZM^2} \Lambda_0^{-M} \sum_Q f(Q) \, \prod_{m=1}^M \prod_{v \in {\rm row}\, m} R_{d_{\bm v}}^{u_{\bm v}}(l_{\bm v},r_{\bm v}).
\end{aligned}
\end{equation}
The above expression involves quantities which, just by themselves, are not stable in the thermodynamic limit ({\it i.e.} they don't have a limit). An important observation is that, among other things, Eq.~\eqref{Eq:Con1} says that the eigenvalue $\Lambda_0$ depends on the lateral size of the lattice such that the quantity $\Lambda_0^\frac{1}{N}$ has a well defined limit as $N \to \infty$. Since it is important to work with quantities that are stable in the thermodynamic limit, we normalize the transfer matrix as
\begin{equation}\label{Eq:ReScale1}
\TM \mapsto \widetilde \TM = \Lambda_0^{-1} \TM = \Lambda_0^\frac{1}{N}\RM_{01} \ldots \Lambda_0^\frac{1}{N}\RM_{0N}.
\end{equation}
such that the re-scaled transfer matrix has the largest eigenvalue equal to one. As one can see from Eq.~\eqref{Eq:ReScale1}, this amounts to re-scaling the $\RM$ matrices which is equivalent to re-scaling the $R_{d_{\bm v}}^{u_{\bm v}}(l_{\bm v},r_{\bm v})$ input by the same amount $\Lambda_0^\frac{1}{N}$, which is stable in the thermodynamic limit. The re-scaled transfer matrix will have a stable spectral decomposition
\begin{equation}
\widetilde \TM = \sum_{j=0}^{2^{N+1}-1} \lambda_j \, P_j, \quad \lambda_j = \Lambda_j/\Lambda_0.
\end{equation}
In particular, note that $\lambda_0 =1$. In the limit $N \to \infty$, the spectrum of $\widetilde \TM$ below 1 becomes denser and denser until it degenerates into continuum spectrum. As such, $\lambda_1$ is not isolated, in general, hence it is useful to think of $\lambda_1$ as the edge of the spectrum below $1$.

With this re-scaling,
\begin{equation}\label{Eq:Exp2}
\begin{aligned}
\EM(f) = \lim_{\Ll \to \ZM^2}  \sum_Q f(Q) \, \prod_{m=1}^M \prod_{v \in {\rm row}\, m} \widetilde R_{d_{\bm v}}^{u_{\bm v}}(l_{\bm v},r_{\bm v}),
\end{aligned}
\end{equation}
and the gain here is that each entry in the above expression is stable in the thermodynamic limit. Furthermore, the computation of Eq.~\eqref{Eq:Exp2} can proceed as
\begin{equation}\label{Eq:Exp3}
\begin{aligned}
\EM(f) & = \lim_{\Ll \to \ZM^2}  \sum_Q \, \prod_{m=1}^K \prod_{v \in {\rm row}\, m} \widetilde R_{d_{\bm v}}^{u_{\bm v}}(l_{\bm v},r_{\bm v}) \\
& \times f(Q) \, \prod_{m=K+1}^P \prod_{v \in {\rm row}\, m} \widetilde R_{d_{\bm v}}^{u_{\bm v}}(l_{\bm v},r_{\bm v}) \\
& \times \prod_{m=P+1}^{P+L} \prod_{v \in {\rm row}\, m} \widetilde R_{d_{\bm v}}^{u_{\bm v}}(l_{\bm v},r_{\bm v}),
\end{aligned}
\end{equation}
where the product in the second line covers the rows that intersect with the domain $\Dd$. Lastly, each line can be calculated along the lines described in section~\ref{Sec:TensorCalc}, with the result
\begin{equation}
\EM(f) = \lim_{K,L \to \infty} \langle \Psi | \widetilde \TM^K \Gamma_{\Dd} \widetilde \TM^L | \Psi' \rangle.
\end{equation} 
The matrix $\Gamma_{\Dd}$ is to be computed from the particular expression of physical observable $f(Q)$. However, this matrix is not needed here. Now, to avoid complications related to the boundary conditions, we can simply assume periodic conditions also in the vertical direction, in which case
\begin{equation}
\EM(f) = \lim_{K,L \to \infty} {\rm Tr} \big [P_0 \Gamma_{\Dd} P_0 \big ]=\lim_{K,L \to \infty} {\rm Tr} \big [\Gamma_{\Dd} P_0 \big ],
\end{equation} 
where the asymptotic behaviors of the powers have been used. At this point, we have identified $P_0$ as one of the fundamental object worth of computing. 

\subsection{Correlation functions}
\label{Sub-sec:CorrF}

We consider here a similar physical observable $f_c(Q)$ as before and, this time, we use the index $c$ to indicate that it depends only on bonds close to the center of the lattice. By $f_{c+y}$, we denote the vertical translation by $y$ rows of this observable. Then, one is often interested in mapping the expected value
\begin{equation}
\EM(f_c\cdot f_{c+y}) : = \sum_{Q} f_c(Q) f_{c+y}(Q) \PM_\beta(Q)
\end{equation}
as function of $y$. Following the same arguments as in the previous subsection, it is easy to see that such quantity can be computed as 
\begin{equation}
\EM(f_c\cdot f_{c+y}) : = \lim_{K,L \to \infty} {\rm Tr}\big [\widetilde \TM^K \Gamma_c \TM^{y-1} \Gamma_{c+y} \widetilde \TM^L \big ],
\end{equation}
Using the asymptotic behaviors of the powers, we can write
\begin{equation}
\begin{aligned}
\EM(f_c\cdot f_{c+y})  \approx & {\rm Tr}\big [ P_0 \Gamma_c P_0 \Gamma_{c+y} P_0 \big ] \\
& \qquad + \lambda_1^{y} {\rm Tr}\big [ P_0 \Gamma_c P_1 \Gamma_{c+y} P_0 \big ] ,
\end{aligned}
\end{equation}
in the limit $y \to \infty$. The important conclusion is that
\begin{equation}
\EM(f_c \cdot f_{c+y})  \approx \EM(f_c)\EM(f_{c+y}) + \lambda_1^{y} C.
\end{equation}
At this point we have identified another important quantity, specifically, $\lambda_1$, which determines the asymptotic behavior of the correlation functions with respect to the vertical separation. Let us also note that, in a translational invariant vertex model like the one considered here, $\EM(f_y) = \EM(f_0)$. Furthermore, to derive similar statements for the horizontal direction, one can simply rotate the system by $90^\circ$ and recompute the transfer matrix and its eigenvalue $\lambda_1$. 

\section{Quantum circuit implementation}
\label{Sec:QC}

The identity in Eq.~\eqref{Eq:MatElement} seems to suggest that the partition function of the physical system, with various boundary conditions, can be coded as quantum circuits and evaluated on a quantum computer using $N+1$ qubits. This, however is not exactly the case because of inherent complications spurring from the non-unital character of the transfer matrix $\TM$. Nevertheless, we will show that the action of $\TM$ on arbitrary many-qubit state can be code with quantum circuits and resolved by quantum simulators, at least. This is a step stone for the spectral analysis of $\TM$ \cite{SaadBook}, which eventually will deliver the quantities relevant for the convergence rate and asymptotic behavior of the correlation functions, and much more.

The goal of this section is to present the generic quantum circuits and to pin-point the constraints and the limitations, hereby to ultimately state what exactly will the quantum circuits deliver.
 
\subsection{The generic structure of the quantum circuit}

As we already acknowledge above, we are interested in a circuit with a global structure
\begin{equation}\label{Eq: T's together}
\begin{aligned}
   \underbrace{
	 \Qcircuit @C=1.0em @R=1.0em @!R {
		\lstick{ {q}_{0} :  } & \multigate{4}{\TM} & \multigate{4}{\TM}  & \hdots & \space & \hdots & \space & \multigate{4}{\TM} & \qw\\
	 	\lstick{ {q}_{1} :  } & 	\ghost{\TM} & \ghost{\TM}       			& \hdots & \space & \hdots & \space & \ghost{\TM}  & \qw\\
		\lstick{ {q}_{3} :  } & \ghost{\TM} & \ghost{\TM} 				& \hdots & \space & \hdots & \space & \ghost{\TM}  & \qw\\
		\\
		\lstick{ {q}_{N} :  } & \ghost{\TM} & \ghost{\TM} 				& \hdots & \space & \hdots & \space & \ghost{\TM}  & \qw\\
	} }_\text{M -Times}
\end{aligned}
\end{equation}
The matrix $\TM$ is a product of $\mathbb R_{0k}$ matrices acting on specific pairs of qubits. Therefore, if $\RM$ is the 2-qubit gate
\begin{equation}\label{Eq:RMat}
\mathbb R = R_{d}^{u}(l,r) \, E_{l}^{r} \otimes E_{d}^{u},
\end{equation}
the quantum circuit for $\TM$ takes the form 
\begin{equation}\label{Eq:TMCirc}
\begin{aligned}
{\small
    \Qcircuit @C=1.0em @R=0.3em {
	 	 &\multigate{4}{\TM} & \qw  & \space& \space & \qw & \qw & \qw & \qw & \qw & \qw & \multigate{4}{\begin{matrix} \ \\ \ \\  \  \\ \RM \end{matrix}} & \qw \\
	 	 &\ghost{\TM} & \qw & \space& \space & \qw & \qw & \qw & \qw & \multigate{3}{\begin{matrix} \ \\ \ \\ \RM \end{matrix}} & \qw & \qw & \qw \\
	 	 &\ghost{\TM} & \qw  & =& \space &\qw & \qw & \multigate{2}{\begin{matrix} \ \\ \RM \end{matrix}} & \qw & \qw & \qw & \qw & \qw \\
	 	 &\ghost{\TM} & \qw  & \space& \space &\multigate{1}{ \RM} & \qw & \qw & \qw & \qw & \qw & \qw & \qw \\
	 	 &\ghost{\TM} & \qw  & \space& \space &\ghost{\RM} & \qw & \ghost{\RM} & \qw & \ghost{\RM} & \qw & \ghost{\RM} & \qw \\ 
	 }}
\end{aligned}
\end{equation}

Above, $\TM$ is exemplified for $N=4$, which requires a circuit with five qubits. As advertised in our introduction, the number of qubits and the depth of the circuit grow linearly with the lateral size of the lattice. Even though the gates do not act on adjacent qubits and swap operations could be introduced by the transpiling, depending on the topology of the back end machine, the dependence of the circuit depth on the lattice size still remains linear. This can be seen directly by examining the circuit~\eqref{Eq:TMCirc}, but we have also verified this statement for up to $N=50$, using the {\it transpile} function from the Qiskit library with the basis gates 'Id', 'U' and 'CX'.

\subsection{The matrix $\RM$}
\label{Sub-Sec:RMat}

We express the $\RM$-matrix in the 2-qubit basis $|00\rangle$, $|01\rangle$, $|10\rangle$ and $|11\rangle$, in this order, such that $\RM$ becomes an explicit $4 \times 4$ matrix. We recall that the entries of this matrix can be found in Eq.~\eqref{Eq:RCoeff}. From definition~\eqref{Eq:RMat} and identity~\eqref{Eq:ProjIdentity}, one can see that $R_{d}^{u}(l,r)$ sits at position $(i,j)$ of this $4 \times 4$ matrix, with $i = 2\cdot l + d$ and $j=2\cdot r + u$.

In general, the $\RM$ matrix will not be unitary and not even hermitean. Nevertheless, $\RM$ always accepts a singular value decomposition
\begin{equation}\label{Eq:SVD}
\RM = U \, D \, V,
\end{equation}
where $U$ and $V$ are unitary matrices and $D$ is a diagonal $4 \times 4$ matrix. We write $D$ as ${\rm Diag}(d_0,d_1,d_2,d_3)$ and we note that those diagonal entries are always real and positive numbers. Furthermore, the standard singular value decomposition algorithms always order these entries such that $d_0$ is the largest, hence, this will be assumed from now on. We will re-scale $\RM$ by its largest singular value $d_0$, $\RM \mapsto d_0^{-1} \, \RM$, such that the first entry of $D$ is equal to one and all the other entries are less or equal to one. Correspondingly, we re-scale the matrix $\TM$ as $\TM \mapsto d_0^{-N} \, \widetilde \TM$. As we shall see below, this will make no difference.

\begin{figure*}[t!]
\center
\includegraphics[width=\textwidth]{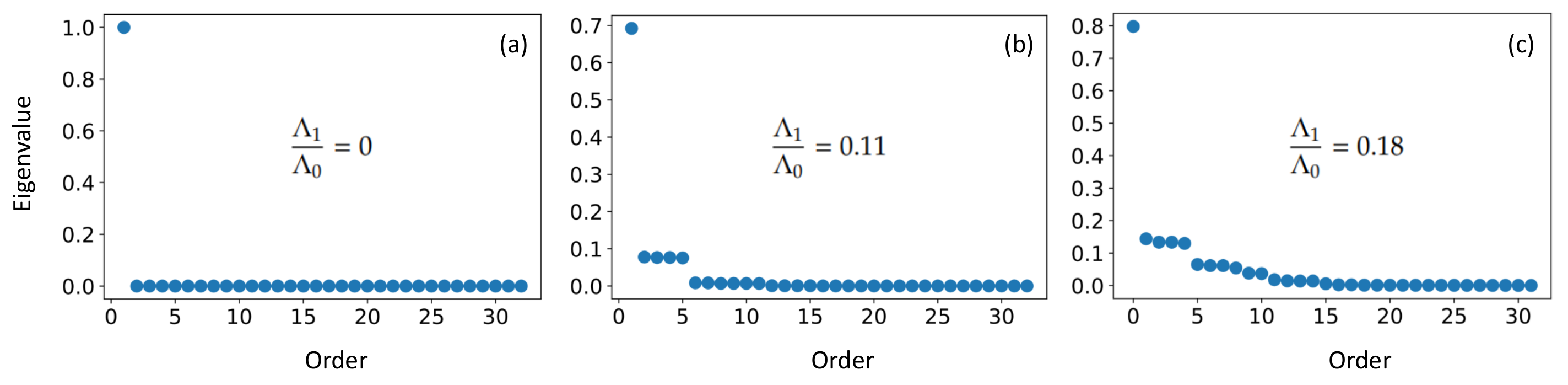}\\
  \caption{The spectra of $\TM$ matrices generated with input data corresponding to (a) $c=10$, (b) $c=0.4$ and (c) $c=0.0$ (right). The corresponding ratios $\lambda_1=\Lambda_1/\Lambda_1$ are displayed in each panel. The $\TM$ matrices were generated with four $\RM$ matrices.}
 \label{Fig:TEig}
\end{figure*} 

As it is well known \cite{TerashimaIJQI2005}, the non-unitary diagonal matrix 
\begin{equation}
D = {\rm Diag}(d_0,d_1,d_2,d_3) = {\rm Diag}(\vec d\,)
\end{equation}
can be implemented by a quantum circuit $\Cc_D$ if one uses one acilla qubit $|\cdot \rangle_a$ and performs projective measurements on the acilla. This will be amply discussed in the next section, where two solutions will be presented, the one from Ref.~\cite{TerashimaIJQI2005} and another one that is optimized for the problem at hand. Here, we only want to specify the actual action of the circuit,
\begin{equation}\label{Eq:CD}
\begin{aligned}
& \Cc_D \, \big (\alpha_0 |00\rangle + \cdots + \alpha_3 |11\rangle \big ) \otimes |0\rangle_a \\
& \qquad \quad = \Nn(\vec d,\vec \alpha) \,  \big (d_0\alpha_0 |00\rangle + \cdots + d_3\alpha_3 |11\rangle \big ) \otimes |0\rangle_a.
\end{aligned}
\end{equation}
If we look at the block inside the parenthesis, the circuit delivers what is needed, but it also inserts the constant
\begin{equation}\label{Eq:Nn}
\Nn(\vec d,\vec \alpha) = 1/ \sqrt{(\vec d \cdot \vec \alpha)^2 }
\end{equation} 
that keeps the 3-qubit state properly normalized. 

At this point we identify one limitation, namely, that a quantum circuit can deliver the action of a diagonal matrix on qubit states only up to a multiplicative constant. This is unavoidable and its consequences are analyzed next.

\subsection{The $\TM$ matrix}
\label{Sub-Sec:TM}

Given the observations made in the previous sub-section, the proposed quantum circuit~\eqref{Eq:TMCirc} has the following action on a generic many-qubit state,
\begin{equation}
\begin{aligned}
\Cc_{\TM} |\Psi\rangle & =  (\Nn_1\, \RM_{01}) \cdots (\Nn_N \, \RM_{0N}) |\Psi \rangle \\ 
& = \Nn_1 \cdots \Nn_N \, \TM|\Psi \rangle.
\end{aligned}
\end{equation} 
As one can see, the proposed quantum circuit delivers the matrix $\TM$, but only up to a multiplicative constant,
\begin{equation}
\Cc_\TM |\Psi \rangle = \Nn_\Psi \TM |\Psi \rangle, \quad \Nn_\Psi = \Nn_1 \cdots \Nn_N.
\end{equation}
As the notation suggests, $\Nn_\Psi$ depends on the many-qubit state. In fact, $\Nn_\Psi$ is precisely the factor that assures the normalization of the state, hence
\begin{equation}
\Nn_\Psi = \frac{1}{\sqrt{\langle \Psi | \TM^\dagger \TM |\Psi \rangle }},
\end{equation}
up to a phase factor. We note that the action of the circuit is not linear with respect to the input.

If $|\tilde \Psi_j^R\rangle$ are the right eigen-vectors normalized such $\langle \tilde \Psi_j^R|\tilde \Psi_j^R\rangle =1$, then
\begin{equation}
\Cc_\TM |\tilde \Psi_j^R \rangle = \Nn_{\tilde \Psi_j^R} \Lambda_j |\tilde \Psi_j^R \rangle
\end{equation}
and, since the right hand is a normalized state, we must conclude that
\begin{equation}
|\Nn_{\tilde \Psi_j^R}| =| \Lambda_j|^{-1}.
\end{equation}
As a consequence, there is no chance to compute with $\Cc_\TM$ any of the eigenvalues, in particular, $\Lambda_0$. However, the other quantities singled out in section~\ref{Sec:PartFunc} are within reach. Indeed, using the asymptotic behavior of the powers, for large $M$'s, we have
\begin{equation}\label{Eq:T12}
\begin{aligned}
(\Cc_{\TM})^M |\Psi\rangle & \approx \Nn_\Psi \Nn_{\TM \Psi} \cdots \Nn_{\TM^{M-1}\Psi} (\Lambda_{0})^M \, P_{\rm 0}|\Psi \rangle \\
 & = \Nn_\Psi (\Lambda_{0})^M \, \alpha_{0}|\Psi_{0}^R \rangle
\end{aligned}
\end{equation}
and, since the result is a normalized many-qubit state,
\begin{equation}\label{Eq:Limit11}
(\Cc_{\TM})^M |\Psi\rangle \rightarrow |\tilde \Psi_{0}^R \rangle \quad {\rm as} \quad M \to \infty,
\end{equation}
for any input many-qubit state $\Psi$ that has a non-zero overlap with $\Psi_{0}^R$. 

As we shall see, this is the case for states with real and positive coefficients. This is an important observation because, since all entries of $\TM$ are real and positive numbers, $\Psi_{0}^R$ has at its turn only real coefficients in the qubit base expansion. This is a feature that makes vertex models special because $\Psi_{0}^R$ can be read off directly from the histograms returned by the quantum simulator. Hence, we can avoid expensive state tomography procedures \cite{DArianoAIEP2003}.

The important conclusion is that we have a practical way to map the right eigen-vector of $\TM$ corresponding to its largest eigenvalue. However, we lack the means to map the eigenvalue itself. Nevertheless, in section~\ref{Sec:Results} we will present a procedure and results on the computation of the ratio $\lambda_1=\Lambda_1/\Lambda_0$.

\section{Qiskit Implementation}
\label{Sec:Qiskit}

Qiskit \cite{Qiskit} is an open source library of Python functions and methods, which enables one to build, analyze and optimize quantum circuits, as well as interface the algorithms with genuine quantum computers or quantum simulators. In this section, we take advantage of the many features offered by Qiskit and develop the full quantum circuit $\Cc_\TM$ that implements the matrix $\TM$, up to a multiplicative constant. The full Python code is appended at the end of the manuscript.

\begin{figure*}[t]
\center
\includegraphics[width=\textwidth]{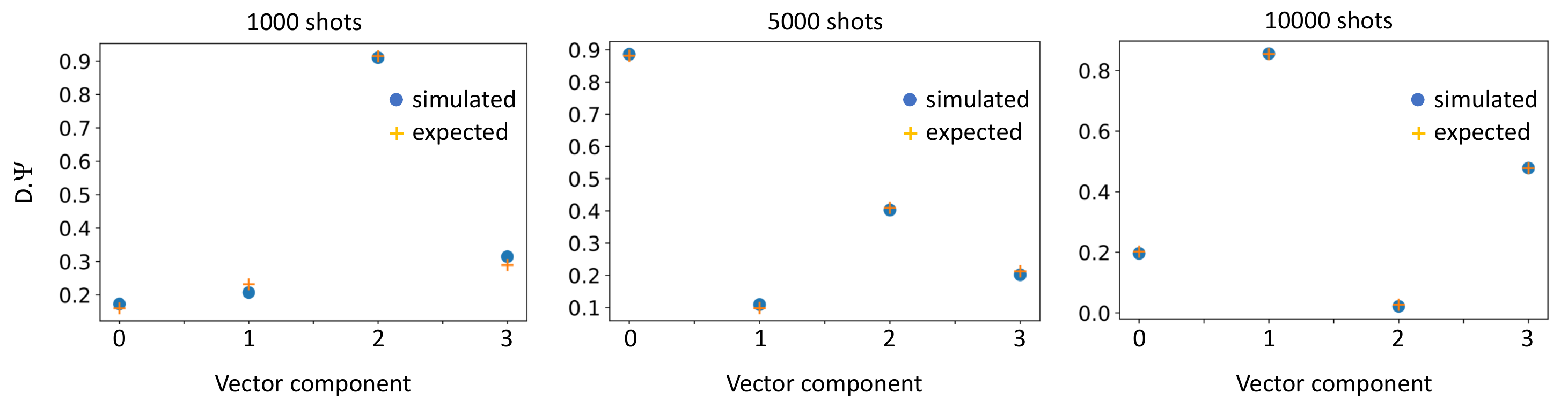}\\
  \caption{The simulated action of a $4\times 4$ diagonal matrix $D$ acting on a normalized 2-qubit state $V$, both generated via a pseudo-random process. The Qasm simulated action is compared with the expected action for an increasing number of shots. In all panels, the number of meaningful runs were more than a half of the total number of shots.}
 \label{Fig:DTest}
\end{figure*}

\subsection{Initialization and input generation}

For the purpose of testing and exemplification of the code functionalities, we need to decide first on a model for the bond energies $\epsilon_d^u(l,r)$. At first, we declare that each active bond contributes with an energy $\Delta$, hence,
\begin{equation}
\epsilon_d^u(l,r) = (d+u+l+r)\Delta.
\end{equation}
This choice leads to a very particular $\RM$ matrix, which has only one non-zero singular value. Likewise, the $\TM$ matrix has only one eigenvalue with the corresponding eigen-vector being supplied by the many-qubit state $|00\ldots 0 \rangle$. In order to sample more general cases, we include in our model a random perturbation:
\begin{equation}
\epsilon_d^u(l,r) = c \cdot (d+u+l+r)\Delta + {\rm Random}() \Delta.
\end{equation}
The parameter $c$, above, enables us to adjust the strength of the non-fluctuating part of the model. Throughout, the unit of energy will be fixed such that $\Delta=1$.

Below is an example of $\RM$ matrix generated with $c=0.4$:
\begin{equation}\label{Eq:RMnum}
\RM = {\scriptsize \begin{bmatrix}
0.5265 & 0.1508 & 0.0963 & 0.0305 \\
0.1941 & 0.1467 & 0.0410 & 0.0370 \\
0.3334 &  0.2018 & 0.1079 & 0.0126 \\
0.1588 & 0.0160 & 0.0546 & 0.0302
\end{bmatrix}.}
\end{equation}
This matrix was generated with $\beta =2$ using the following Python script:
{\scriptsize
\begin{lstlisting}[language=Python,caption={ \ },label={Script:1}]
# prepares the input R
import numpy as np
beta=2.0
fact=0.4
fill="0000"
R=np.zeros((4,4)), #this stores R
eps=[]  #this stores the energies of the bond configs
for i in range(16):
    b=np.base_repr(i,base=2)
    q=len(b)
    b=fill+b
    b=b[q:]
    s=0.0
    for j in b:
        s=s+int(j)
    eps.append(fact*s+np.random.rand())
    u1=2*int(b[0])+int(b[1])
    u2=2*int(b[2])+int(b[3])
    R[u1][u2]=np.exp(-beta*eps[i])
\end{lstlisting}
}

The $\RM$ matrix from Eq.~\eqref{Eq:RMnum} will be used as the input for the tests presented in the following sections. As such, it is useful to have its explicit singular value decomposition:
\begin{equation}\label{Eq:SVD}
\begin{aligned}
U & = {\scriptsize
\begin{bmatrix}
-0.7464  & 0.5215 & 0.3776 & -0.1678\\
-0.3242 &-0.5346  & 0.3963  &0.6722\\
-0.5380 & -0.5525 & -0.4841 & -0.4133\\
-0.2196  & 0.3699 & -0.6825   & 0.5907
\end{bmatrix}} \\
D & = {\scriptsize
\begin{bmatrix} 
1 & 0   &      0    &     0        \\
0    &     0.1553 & 0    &     0        \\
0    &     0    &     0.0511 & 0        \\
0    &     0     &    0     &    0.0437
\end{bmatrix} } \\
V & ={\scriptsize
\begin{bmatrix}
-0.9039 & -0.3672 & -0.2093 & -0.0650 \\
0.3935 & -0.9141 & -0.0968 &  0.0030 \\
0.1570 &  0.1705 & -0.9726 & -0.0142 \\
-0.0578 & -0.0187 & -0.0272 &  0.9977 
\end{bmatrix} }
\end{aligned}
\end{equation}
Above, the singular values have been already normalized. For reader's convenience we supply the Python script which produced the singular value decomposition:
{\scriptsize
\begin{lstlisting}[language=Python,caption={ \ },label={Script:1}]
#executes the singular value decomposition
U,d,V=np.linalg.svd(R)
d=d/d[0] # normalizes the singular values
\end{lstlisting}
}

As we already stressed in sub-section~\ref{Sub-Sec:TM}, the spectrum of the $\TM$ matrix dictates the rate of convergence towards the thermodynamic limit. Samples of such spectra for $\TM$ matrices generated with four $\RM$ matrices (hence $N=4$) are shown in Fig.~\ref{Fig:TEig}. In panel (a), the $\TM$ matrix was generated with $c=10$, hence the random component is insignificant and the spectrum contains one single non-zero eigenvalue. The spectrum in panel (b) corresponds to the $\TM$ matrix generated with the $\RM$ matrix from Eq.~\eqref{Eq:RMnum}, which we recall that it was generated with $c=0.4$, hence it has a moderate random component. In panel (c), the $\TM$ matrix was generated with $c=0$, hence entirely by a random process. In all cases, the first eigenvalue is significantly larger then the rest of the eigenvalues, which will assure a rapid convergence towards the thermodynamic limit. This seems to be a feature of our model.

Qiskit enables the insertion of any unitary matrix in a quantum circuit. The following line of code exemplifies the process of appending to an existing circuit a unitary operator $W$ acting on two qubits $i$ and $j$:
{\scriptsize
\begin{lstlisting}[language=Python,caption={ \ },label={Script:1}]
# append a two-qubit unitary gate W
circuit.unitary(W,[i,j]) 
\end{lstlisting}
}
The unitary gates inserted in this fashion will be transformed into elementary gates when the final circuit is being transpiled. Hence, given the singular value decomposition of the $\RM$ matrix and the above observations, the remaining challenge is the implementation of the diagonal matrix D.

\begin{figure*}[t!]
\center
\includegraphics[width=\textwidth]{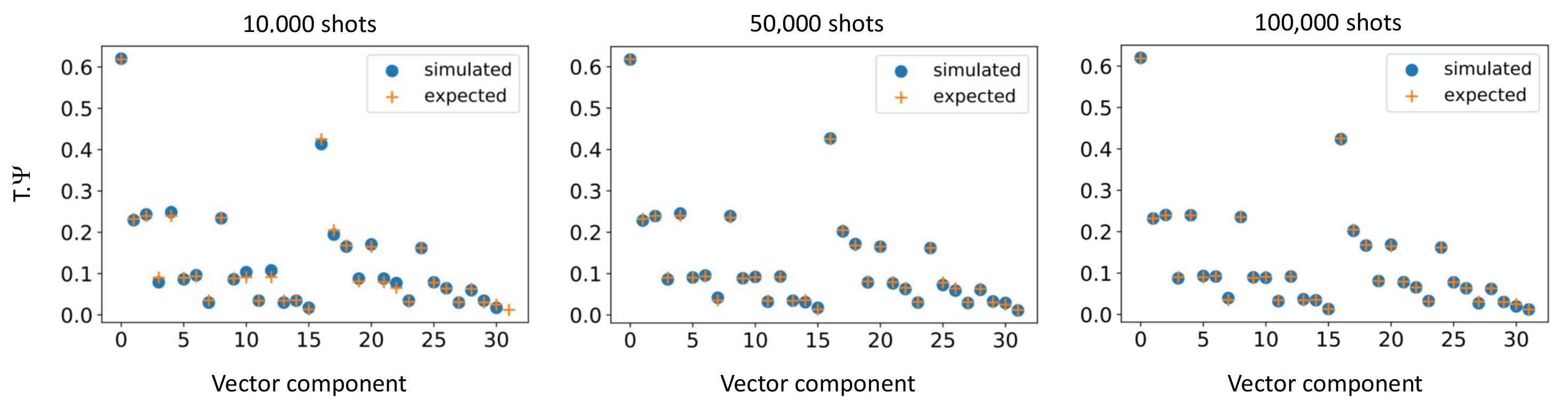}\\
  \caption{The simulated action of the $\TM$ matrix the $5$-qubit state $|\Psi\rangle = |00\ldots0\rangle$, using the circuit~\eqref{Eq:FinalT} and the $\RM$ matrix~\eqref{Eq:RMnum}. The Qasm simulated action is compared with the expected action for an increasing number of shots.}
 \label{Fig:TP1test}
\end{figure*}

\subsection{A quantum circuit for generic diagonal matrix}

We review here a quantum circuit proposed in \cite{TerashimaIJQI2005} for implementing a diagonal matrix, which served as a model for the actual quantum circuit used in our work.
  
We first recall that the singular values are scaled such that $D$ takes the form
\begin{equation}
D = {\rm Diag}(1,d_1,d_2,d_3), \quad d_i <1,
\end{equation}
and such matrix can be always decomposed as
\begin{equation}\label{Eq:Deco}
D = {\rm Diag}(1,d_1,1,1) {\rm Diag}(1,1,d_2,1) {\rm Diag}(1,1,1,d_3).
\end{equation}
Furthermore,
\begin{equation}\label{Eq: D's decomposed}
\begin{aligned} 
{\rm Diag}(1,d_1,1,1)&= (X \otimes I)N(d_1)(X \otimes I)\\
{\rm Diag}(1,1,d_2,1)&= (I \otimes X)N(d_2)(I \otimes X)\\
{\rm Diag}(1,1,1,d_3)&= (I \otimes I)N(d_3)(I \otimes I)\\
\end{aligned}
\end{equation}
where
\begin{equation}\label{N(a) matrix}
N(a) :={\scriptsize
\begin{bmatrix}
1&0&0&0\\
0&1&0&0\\
0&0&1&0\\
0&0&0&a
\end{bmatrix}}.
\end{equation}
At this point one arrives at the following quantum circuit for the matrix $D$ \cite{TerashimaIJQI2005}:
\begin{equation}\label{first N(a) circuit}
\begin{aligned}
{\small
    \Qcircuit @C=0.8em @R=0.5em @!R {
	 	\lstick{}  & \qw & \qw   & \ctrl{1} & \qw & \gate{X}  & \qw & \ctrl{1}  & \qw & \gate{X} & \qw& \ctrl{1} & \qw\\
	 	\lstick{}   & \gate{X} & \qw & \gate{N(d_1)} & \qw & \gate{X} & \qw & \gate{N(d_2)} & \qw & \qw & \qw & \gate{N(d_3)} & \qw 
	 }}
	 \end{aligned}
\end{equation}

In \cite{TerashimaIJQI2005}, the action of $N(a)$ is generated with the help of an ancilla qubit and a projective measurement as it follows: 
\begin{equation}\label{Eq:Na}
\begin{aligned}
{\small
    \Qcircuit @C=1.0em @R=0.0em @!R {
\lstick{} & \gate{N(a)} & \qw & \qw & \lstick{}\\
 }
=
    \Qcircuit @C=1.0em @R=0.0em @!R {
\lstick{} & \lstick{} & \ctrl{1} & \qw & \qw & \qw\\
	 	\lstick{} &\lstick{|0\rangle} & \gate{S(a)} & \gate{P(0)} & \qw & \qw & \lstick{}& |0\rangle\\
 }
 }
 \end{aligned}
\end{equation}
In diagram~\eqref{Eq:Na}, it can be seen that the main qubit is used as control node and the ancilla qubit starts in state zero and is acted by two operations, $S(a)$ and $P(0)$. $S(a)$ is a unitary gate corresponding to the following matrix 
\begin{equation}
\begin{aligned} 
S(a)=
\begin{bmatrix}
a& \sqrt{1-a^2}\\
\sqrt{1-a^2}&-a
\end{bmatrix},
\end{aligned}
\end{equation}
and, since $S(a)$ is unitary, Qiskit is able to transpile it to elementary gates, as we already mentioned for the unitary matrices obtained from the single value decomposition. However, the other operation is 
\begin{equation}
P(0)={\small \begin{bmatrix}1& 0\\0&0\end{bmatrix}},
\end{equation} 
hence it is a projective measurement on the zero state. There are two steps to properly integrate this gate. The first step is to add a measurement after $S(a)$ and record the measurement in a classical register. The second step is to examine the classical register after running the circuit and select only those shots that produced value $0$. 

To conclude, the quantum circuit for implementing the diagonal matrix $D$ proposed in \cite{TerashimaIJQI2005} is:
\begin{equation}\label{Eq:FullD1}
\begin{aligned}
{\scriptsize
    \Qcircuit @C=1.0em @R=0.0em @!R {
	 	\lstick{q_0:}  & \qw & \ctrl{1} & \qw \barrier[0em]{2} & \qw & \gate{X} & \ctrl{1} & \qw \barrier[0em]{2} & \qw & \gate{X} & \ctrl{1} & \qw & \qw \\
	 	\lstick{q_1:}  & \gate{X} & \ctrl{1} & \qw & \qw & \gate{X} & \ctrl{1} & \qw & \qw & \qw & \ctrl{1} & \qw & \qw \\
	 	\lstick{a_{\ }:} & \qw & \gate{S(d_1)} & \meter & \qw & \qw & \gate{S(d_2)} & \meter & \qw & \qw & \gate{S(d_3)} & \meter & \qw \\
	 	\lstick{c_{\ }:} & {/_{_{5}}} \cw & \cw & \dstick{1} \cw \cwx[-1] & \cw  & \cw & \dstick{2} \cw \cwx[-1] & \cw & \cw & \cw & \cw & \dstick{3} \cw \cwx[-1] & \cw
	 }}
	 \end{aligned}
\end{equation}
For reader's convenience, the diagram below shows a step by step progression of the circuit:
\begin{equation}
{
\begin{tabular}{ |c|c|c|c|c|c| } 
\hline
\backslashbox{Action}{State} & $ |000\rangle$ & $|001\rangle$ & $|010\rangle$ & $|011\rangle $ & $|111\rangle $ \\
\hline
start & $ \alpha $ & $\beta$ & $\gamma$ & $\delta$ & 0 \\
$\myatop{I}{X}$ & $ \gamma $ & $\delta$ & $\alpha $ & $\beta$ & 0 \\
$CCS$&$ \gamma$ & $\delta$ & $\alpha $ & $d_1 \beta$ & $\sqrt{1-{d_1}^2}\beta $\\
measurement & $ \gamma $ & $\delta$ & $\alpha $ & $d_1 \beta $ & 0 \\
$\myatop{X}{X}$ & $d_1 \beta $ & $\alpha$ & $\delta$ & $\gamma$ & 0 \\
$CCS$&$ d_1 \beta $ & $\alpha $ & $\delta$ & $d_2\gamma$ & $ \sqrt{1-{d_2}^2}\gamma$\\
measurement & $d_1 \beta$ & $\alpha $ & $\delta$ & $d_2 \gamma $ & 0 \\
$\myatop{X}{I}$ & $\alpha $ & $d_1\beta $ & $d_2\gamma $ & $\delta$ & 0 \\
$CCS$&$ \alpha$ & $d_1 \beta $ & $d_2 \gamma $ & $d_3 \delta$ & $\sqrt{1-{d_3}^2}\delta$ \\
measurement & $\alpha $ & $d_1\beta $ & $d_2 \gamma $ & $d_3 \delta $ & 0 \\ 
\hline
\end{tabular}}
\end{equation}
Above, only the measurements that returned 0 are considered and the normalization of the many-qubit state was omitted.

\begin{figure*}[t!]
\center
\includegraphics[width=\textwidth]{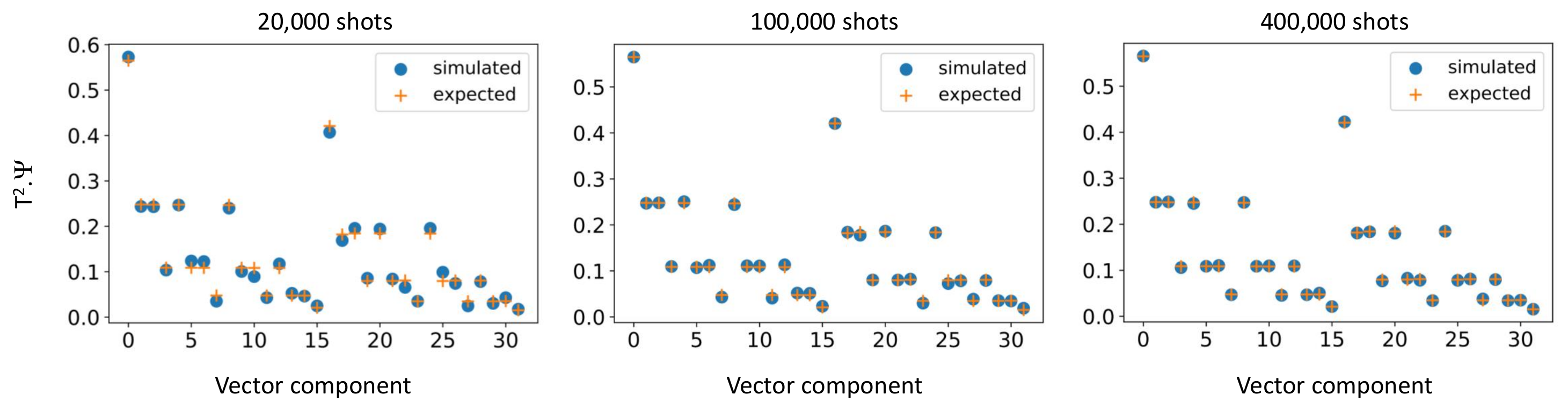}\\
  \caption{Same as Fig.~\ref{Fig:TP1test} but for the action of $\TM^2$ on the same $5$-qubit state.}
 \label{Fig:TP2test}
\end{figure*}

\subsection{An optimized quantum circuit for generic diagonal matrix}

The circuit~\eqref{Eq:FullD1} involves three measurements and the simulation is used only if all these three measurements return the 0 state of the ancilla qubit. Since this circuit is repeated many times when assembled into the full $\Cc_{\TM}$ circuit, this creates a problem because a large number of shots will be discarded possibly leading to poor histograms. Thus, it is imperative to reduce the number of measurements, ideally, to just one. 

This can be achieved by combining the strategy of \cite{TerashimaIJQI2005} with the one proposed in \cite{QIP162017}. Indeed, we can promote the matrix $D$ to the $8\times 8$ unitary matrix
\begin{equation}\label{Eq:DM}
\DM = \begin{bmatrix} D & \sqrt{I_4-D^2} \\ \sqrt{I_4-D^2} & -D \end{bmatrix}
\end{equation}
and view this matrix as a unitary operator acting on the 3-qubit states, including the ancilla, with the states expanded in the standard qubit basis 
\begin{equation}
|i_a i_1 i_0 \rangle = |i_a \cdot 2^2 + i_1 \cdot 2 + i_0 \rangle, \quad i_\alpha \in \{0,1\}.
\end{equation}
The action of this operator on the 3-qubit states can be easily understood if we re-write such states in the following form
\begin{equation}
|\psi\rangle = |0\rangle_a \otimes |\psi_{10}\rangle + |1\rangle_a \otimes |\psi'_{10}\rangle \mapsto \begin{bmatrix} |\psi_{10}\rangle \\ |\psi'_{10}\rangle \end{bmatrix},
\end{equation}
which, of course, can be always achieved in a unique fashion. Then the $2\times 2$ blocked matrix~\eqref{Eq:DM} acts naturally on this last 2-row column matrix. The observation is that, if the ancilla qubit starts in state $|0\rangle_a$, then
\begin{equation}
\DM|\psi_{a10}\rangle = |0\rangle_a \otimes D|\psi_{10}\rangle + |1\rangle_a \otimes \sqrt{1-D^2} |\psi_{10}\rangle.
\end{equation}
As such, a projective measurement of the ancilla qubit with output zero will return the 3-qubit state $|0\rangle_0 \otimes D|\psi_{10}\rangle$, up to the normalization constant~\eqref{Eq:Nn}.

To summarize, our proposed circuit for implementing the action of the diagonal matrix $D$ is shown below
\begin{equation}\label{Eq:FinalD}
\begin{aligned}
\small{
    \Qcircuit @C=1.0em @R=0.0em @!R {
	 	\lstick{ {q}_{0} :  } & \multigate{2}{\DM} & \qw & \qw & \qw\\
	 	\lstick{ {q}_{1} :  } & \ghost{\DM} & \qw & \qw & \qw\\
	 	\lstick{ a_{\ } :  } & \ghost{\DM} & \meter & \qw & \qw\\
	 	\lstick{c_{\ }:} & {/_{_{1}}} \cw & \dstick{{}^0} \cw \cwx[-1] & \cw & \cw\\
	 }}
\end{aligned}
\end{equation}
and, as desired, it involves one single measurement of the ancilla qubit. Note that the matrix $\DM$ from Eq.~\eqref{Eq:DM} is unitary precisely because we scaled its entries to be smaller than 1.

As a test, we used the Qasm simulator and followed the following protocol:
\begin{itemize}
\item The three qubits were initialized into a state $|0\rangle_a \otimes (\alpha_0 |00\rangle + \cdots + \alpha_3 |11\rangle)$, with $\alpha$'s properly normalized, using a pseudo-random process;
\item The diagonal entries of the $D$ matrix were initialized using a pseudo-random process;
\item The unitary circuit $\DM$ was transpiled and executed on the Qasm simulator;
\item All three qubits were measured.
\end{itemize}

The results of the tests are reported in Fig.~\ref{Fig:DTest}. In these experiments, we used a total number of shots that ranges from $10^3$ to $10^4$, such that we can sample situations where the simulations are economical but the matching between the simulated and the expected outcomes is less perfect, to situations where the simulations are more demanding and the matching is much improved. The output from the simulator was processed as follows. From the counts of the occurrences of $|0i_1 i_0\rangle$, we generated a histogram and we compared this histogram with the expected result, which can be read off from Eq.~\ref{Eq:CD}, specifically,
\begin{equation}\label{Eq:Prediction1}
P_{0i_1i_0} = \frac{(d_i \alpha_i)^2}{\Nn(\vec \alpha)^2}, \quad i = 2i_1+i_0.
\end{equation}

\begin{figure*}[t!]
\center
\includegraphics[width=\textwidth]{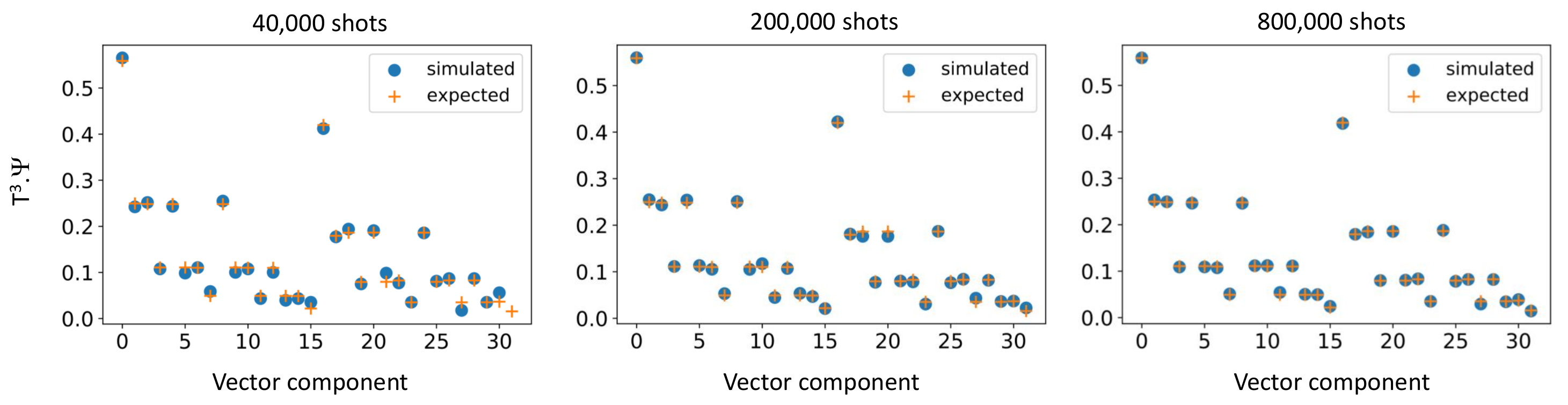}\\
  \caption{Same as Fig.~\ref{Fig:TP1test} but for the action of $\TM^3$ on the same $5$-qubit state.}
 \label{Fig:TP3test}
\end{figure*}

Since the hits for any of the occurrences $|1i_1 i_0\rangle$ have been discarded, the results depend on the number of meaningful shots left to generate the relevant histograms. When the number of meaningful shots was greater than a third of the total number of shots and the latter was $10^4$, we found that the histograms returned by the quantum simulator consistently reproduce the prediction~\eqref{Eq:Prediction1}, exactly up to the second significant digit. The Qiskit code used for the testing we just explained is attached at the end of our manuscript.

\subsection{The complete $\TM$ circuit}

We now have all the components to assemble the full quantum circuit. Below, we unpack the $\TM$ circuit shown in Eq.~\eqref{Eq:TMCirc}:

\begin{widetext}
\begin{equation}\label{Eq:FinalT}
\begin{aligned}
\small{
    \Qcircuit @C=1.0em @R=0.0em @!R {
	 	\lstick{ {q}_{0} :  } & \qw & \qw & \qw & \qw & \qw & \qw & \qw & \qw & \qw & \multigate{4}{V} & \multigate{5}{\DM} & \multigate{4}{U} &  \qw \barrier{5} & \qw & \qw & \qw &\qw & \qw & \meter \\
	 	\lstick{ {q}_{1} :  } & \qw &\qw & \qw & \qw &\qw & \qw & \multigate{3}{V} & \multigate{4}{\DM} & \multigate{3}{U} & \qw & \qw & \qw & \qw & \qw & \qw & \qw & \qw & \meter & \qw \\
	 	\lstick{ {q}_{2} :  } & \qw & \qw & \qw & \multigate{2}{V} & \multigate{3}{\DM} & \multigate{2}{U}  & \qw & \qw & \qw & \qw & \qw & \qw & \qw &\qw & \qw & \qw & \meter & \qw & \qw \\
	 	\lstick{ {q}_{3} :  } & \multigate{1}{V} & \multigate{2}{\DM} & \multigate{1}{U} & \qw & \qw & \qw & \qw & \qw & \qw & \qw & \qw & \qw & \qw & \qw & \qw & \meter & \qw & \qw & \qw \\
	 	\lstick{ {q}_{4} :  } & \ghost{V} & \ghost{\DM} & \ghost{U} & \ghost{V} & \ghost{\DM} & \ghost{U} & \ghost{V} & \ghost{\DM} & \ghost{U} & \ghost{V} & \ghost{\DM} & \ghost{U} & \qw & \qw & \meter & \qw & \qw & \qw & \qw \\ 
	 		 	\lstick{ a_{\ } :  } & \qw & \ghost{\DM} & \meter & \qw & \ghost{\DM} & \meter & \qw & \ghost{\DM} & \meter & \qw & \ghost{\DM} & \meter & \qw & \qw &\qw & \qw & \qw & \qw \\
	 	\lstick{c_{\ }:} & {/_{_{4+5}}} \cw & \cw  & \dstick{{}^8} \cw \cwx[-1]  & \cw\ & \cw & \dstick{{}^7} \cw \cwx[-1] & \cw\ & \cw & \dstick{{}^6} \cw \cwx[-1] & \cw\ & \cw & \dstick{{}^5} \cw \cwx[-1] & \cw & \cw & \dstick{{}^4} \cw \cwx[-2]  & \dstick{{}^3} \cw \cwx[-3] & \dstick{{}^2} \cw \cwx[-4] & \dstick{{}^1} \cw \cwx[-5] & \dstick{{}^0} \cw \cwx[-6]\\
	 }}
\end{aligned}
\end{equation}
\end{widetext}

The quantum circuit corresponds to $N=4$ and, in the following, we specialize the discussion to this particular case. When applied to a many-qubit state $|0\rangle_a \otimes |j_4\ldots j_0\rangle$, the circuit will return the state $|0\rangle_a \otimes \TM |j_4\ldots j_0\rangle$ up to a multiplicative constant $\Nn$, provided the first four digits of the classical register $c$ are 0 at the end of the run. Hence, we have the means to reproduce the state $\Nn \TM |j_4\ldots j_0\rangle$ and to eventually measure the state. Henceforth, we ran the circuit~\ref{Eq:FinalT} a large number of times and access the counts. Next, by selecting the counts for the hits $0000i_4 \ldots i_0$, $i_\alpha \in \{0,1\}$, we made sure that only the meaningful cases were considered, where the many-qubit state was precisely $\Nn|0\rangle_a \otimes \TM |j_4\ldots j_0\rangle$ before measuring the qubits $q_0, \ldots, q_4$. The next step is to generate the probabilities
\begin{equation}
P^{i_4 \ldots i_0}_{j_4 \ldots j_0} = \frac{{\rm counts}(0000i_4 \ldots i_0)}{\sum {\rm counts}(0000i_4 \ldots i_0)}
\end{equation}
and finally to extract the desired information
\begin{equation}
\Nn \langle i_4 \ldots i_0 |\TM| j_4 \ldots j_0 \rangle = \sqrt{P^{i_4 \ldots i_0}_{j_4 \ldots j_0}}.
\end{equation}
The full Qiskit code which generates randomized initial state, builds and runs the circuit as well as it processes the output is appended at the end of our manuscript.

\section{Results}
\label{Sec:Results}

The first part of this section analyzes the performance of the quantum circuits using the Qasm simulator. Our conclusion is that, given enough total shots, the simulated outputs reproduce the expected outputs, up to at least two significant digits.

In the second part, we design quantum simulations to extract physical quantities of interest, such as the convergence rate towards the thermodynamic limit and spectral information about the transfer matrix.

\subsection{Tests and performance}
\label{Sub-sec:Tests}

The simulations of $\TM|\Psi\rangle$, using the circuit~\eqref{Eq:FinalT} and $\RM$ matrix from Eq.~\ref{Eq:RMnum},  are reported in Fig.~\ref{Fig:TP1test}. As before, we varied the number of shots in order to display both economical and more demanding simulations. The largest number of shots was chosen such that the simulated and expected action of $\TM$ match exactly, up to the second significant digit. The expected action of $\TM$ was derived using simple Python scripts involving only the numpy library. The many-qubit state $|\Psi\rangle$ was chosen to be the eigen-vector corresponding to non-zero eigenvalue of the $\TM$ matrix generated with $c=0$, namely, $|00 \ldots 0\rangle$. The reason for this choice is that we expect this vector to have a non-zero overlap with the eigen-vector $\Psi_0^R$ of the simulated $\TM$ matrix. 

\begin{figure}[t!]
\center
\includegraphics[width=0.9\linewidth]{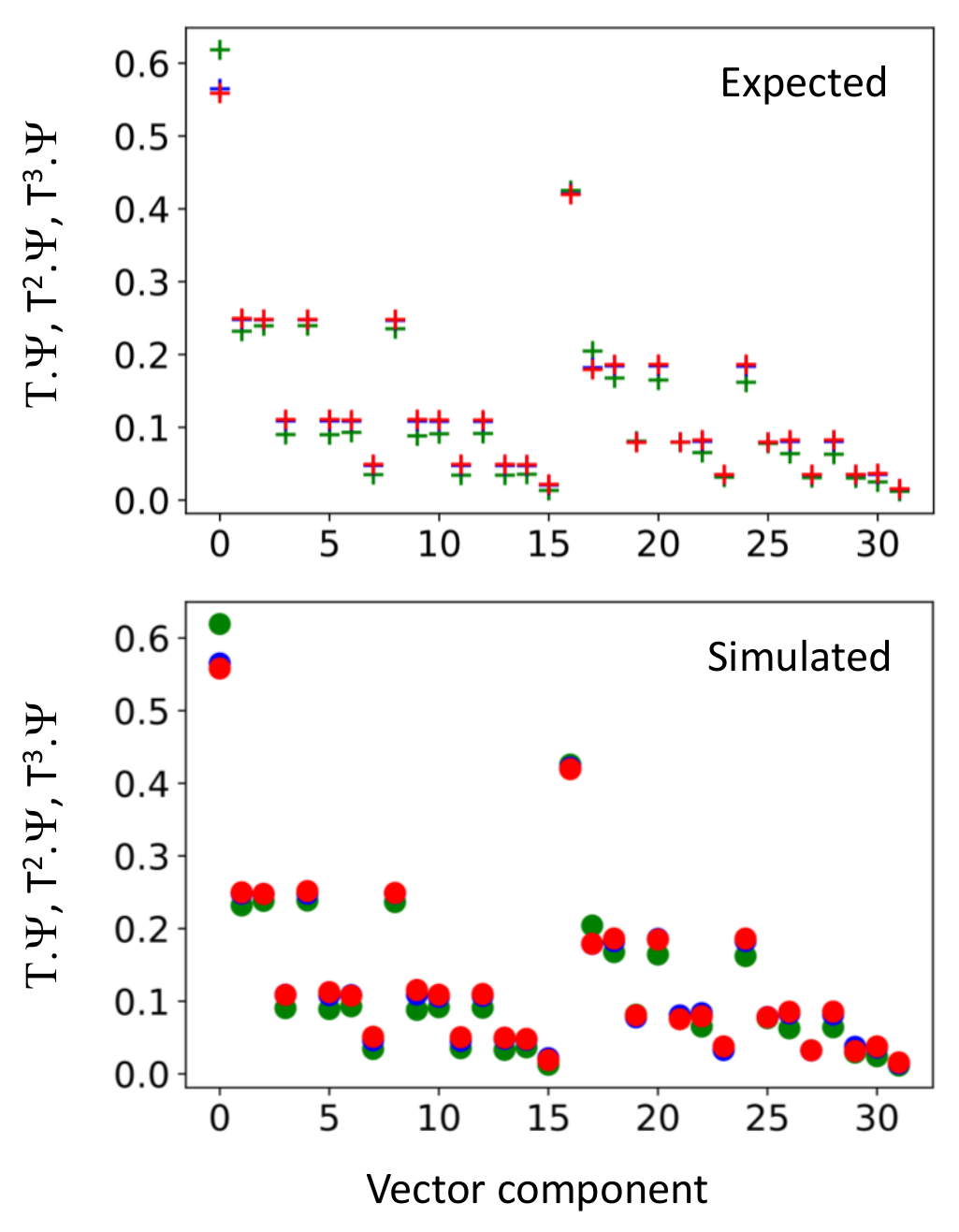}\\
  \caption{Convergence of the expected and simulated actions of $\TM$, $\TM^2$ and $\TM^3$ on the $N$-qubit state $|00\ldots0\rangle$. The Qasm simulated action was generated with the $\RM$ matrix from Eq.~\eqref{Eq:RMnum}, corresponding to a $c=0.4$, and with $800,000$ shots. The spectrum of the corresponding $\TM$ matrix is reported in Fig.~\ref{Fig:TEig}(b) and the ratio $\Lambda_1/\Lambda_0$ is $0.11$.}
 \label{Fig:PowersC0p4}
\end{figure}

Similar tests for the actions $\TM^2|\Psi\rangle$ and $\TM^3|\Psi\rangle$ are reported in Figs.~\ref{Fig:TP2test} and \ref{Fig:TP3test}, respectively. The data in each figure was generated with quantum circuits in which the $\TM$ block was repeated an appropriate number of time. As expected, to achieve the same level of accuracy, the total number of shots needs to be increased with the power of the $\TM$ matrix. For example, to reproduce the first two significant digits of the expected action of $\TM^3$, we need as much as $8\times 10^5$ shots.

An alternative way to implement the action of $\TM^m$ on a many-qubit state is to apply the $\TM$s one at a time. More precisely, compute $|\Psi_1 \rangle=\Cc_{\TM}|\Psi\rangle$ from the histogram and then use the outcome to compute $|\Psi_2\rangle = \Cc_{\TM}|\Psi_1\rangle$, and so on. One difference between the methods is that the  states $|\Psi_1\rangle$, $|\Psi_2\rangle$, etc., are passed in an exact fashion among the $\TM$ blocks, if the first method is used, while they are passed with certain approximation if the second method is used. The first method, which produced the data in Figs.~\ref{Fig:TP2test} and \ref{Fig:TP3test}, can achieve more accurate outputs but one advantage of the second method is the larger percentage of the meaningful shots for each circuit run. For example, to compute $|\tilde \Psi_0^r\rangle$ via the iterative process~\eqref{Eq:Limit11}, the second method is preferable because it can handle large values of $m$ more efficiently. 

\begin{figure}[t]
\center
\includegraphics[width=0.9\linewidth]{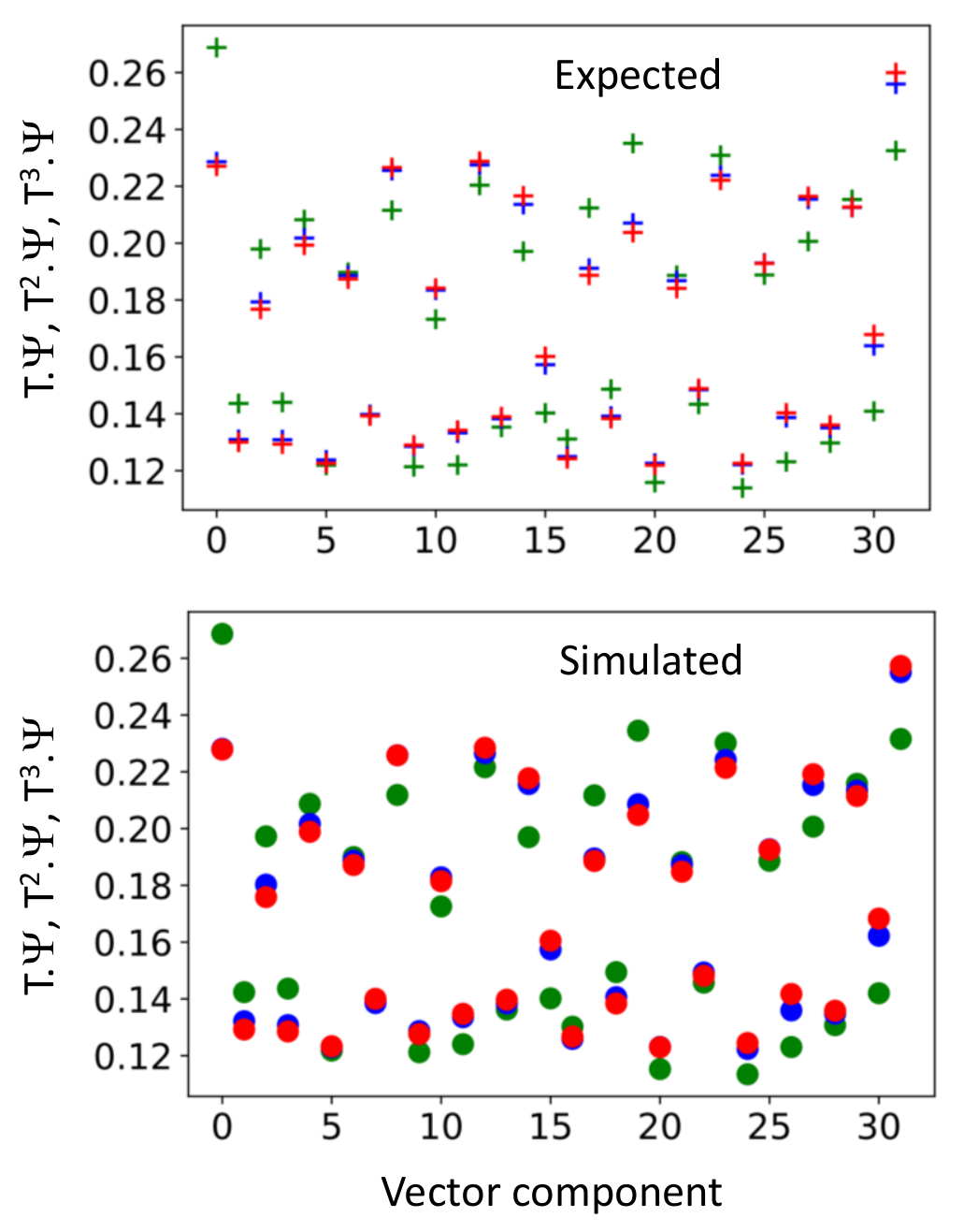}\\
  \caption{Convergence of the expected and simulated actions of $\TM$, $\TM^2$ and $\TM^3$ on a randomized $N$-qubit state. The Qasm simulated action was generated with an $\RM$ matrix corresponding to a $c=0.0$ and with $800,000$ shots. The spectrum of the corresponding $\TM$ matrix is reported in Fig.~\ref{Fig:TEig}(c) and the ratio $\Lambda_1/\Lambda_0$ is $0.18$.}
 \label{Fig:PowersC0p0}
\end{figure}

\subsection{Physical insight}

\begin{figure*}[t!]
\includegraphics[width=\textwidth]{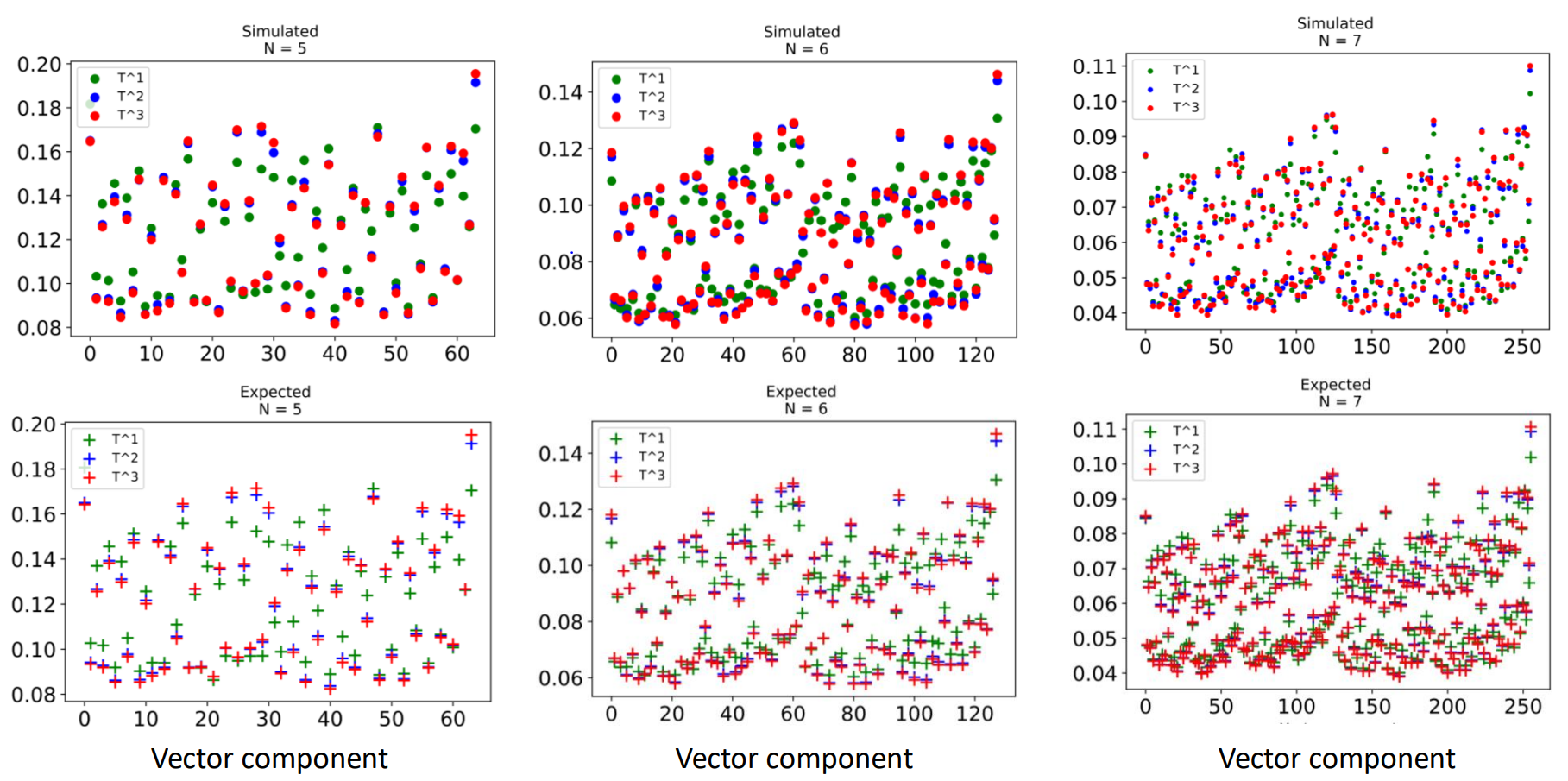}\\
  \caption{Same as Fig.~\ref{Fig:PowersC0p0} but for $N=5$, $6$ and $7$, as indicated by the labels.}
 \label{Fig:NDep}
\end{figure*}

As we already stated in sub-section~\ref{Sub-Sec:TM}, iterated applications of the $\TM$ matrix on any initial many-qubit state will give us access to the normalized eigen-vector $\tilde \Psi_0^R$ corresponding to the largest eigenvalue. With this in mind, we have collected in Fig.~\ref{Fig:PowersC0p4} the expected and the simulated results from our tests reported in Figs.~\ref{Fig:TP1test}, \ref{Fig:TP2test} and \ref{Fig:TP3test}. As one can see, the convergence to the limit stated in Eq.~\ref{Eq:Limit11} is achieved with $M=3$, hence, at least for this particular case, we can indeed resolve the eigen-vector  $\Psi_0^R$, up to two significant digits. The convergence rate for the limit~\ref{Eq:Limit11} is determined by the ratio $\lambda_1=\Lambda_1/\Lambda_0$ and, as reported in Fig.~\ref{Fig:TEig}, its value is $0.11$ for the simulations reported in Fig.~\ref{Fig:PowersC0p4}. As such, the fast convergence witnessed in Fig.~\ref{Fig:PowersC0p4} is expected.

In Fig.~\ref{Fig:TEig}, we report on another case with a ratio $\lambda_1=\Lambda_1/\Lambda_0=0.18$, hence the convergence is expected to be slower in this case. The simulations performed for this case are reported in Fig.~\ref{Fig:PowersC0p0} and they confirm this expectation. Nevertheless, the convergence of the first two significant digits is achieved for $M=4$.

We now recall the discussion from sub-section~\ref{SubSec:ExpV}, where we argued that the ratio $\lambda_1 = \Lambda_1/\Lambda_0$ is stable in the thermodynamic limit. This implies, among other things, that the convergence rate should be stable when $N$ is increased. To confirm this statement, we repeated the simulations from Fig.~\ref{Fig:PowersC0p0} for $N=5$, $6$ and $7$ and we indeed observed the similar convergence rates. The results of the simulations are reported in Fig.~\ref{Fig:NDep}. 

\begin{figure*}[t]
\includegraphics[width=0.9\linewidth]{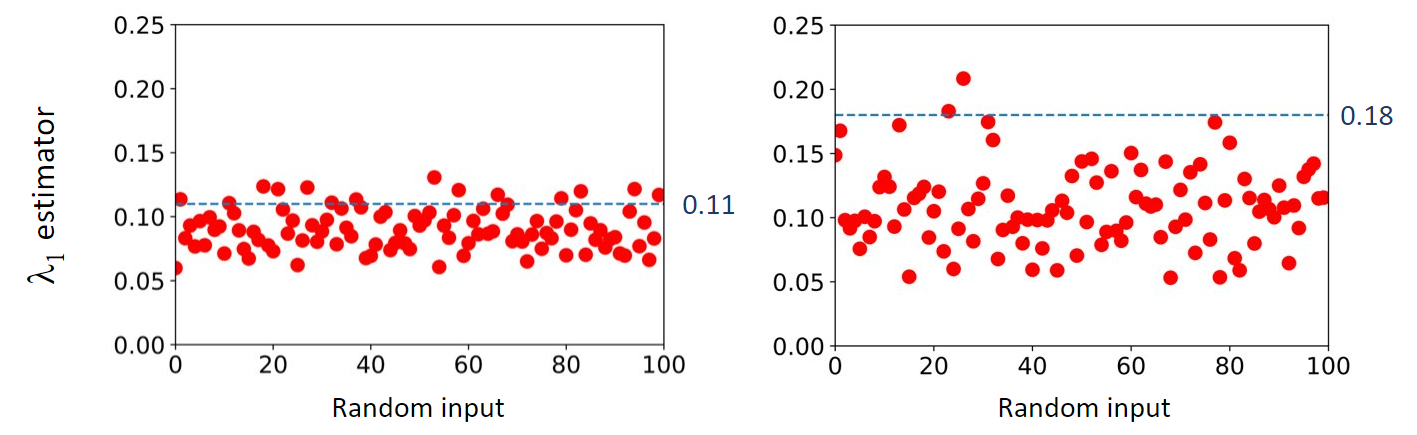}\\
  \caption{Output of the simulated estimator from Eq.~\eqref{Eq:Estimator}, as applied to the $\TM$ matrix analyzed in Fig.~\ref{Fig:PowersC0p4} (left panel) and to the $\TM$ matrix analyzed in Fig.~\ref{Fig:PowersC0p0} (right panel). The markings indicate the expected value of $\lambda_1$. In both cases, the input many-qubit state $|\Psi\rangle$ was sampled 100 times via a pseudo-random process and, for each input, the estimator was evaluated using $10^5$ shots.}
 \label{Fig:Lambda1}
\end{figure*}

We now explain a simple algorithm for estimating the ratio $\lambda_1 = \Lambda_1/\Lambda_0$. The complication here is that the quantum circuit always normalizes the action of $\TM$ on a many-qubit state. Henceforth, we start from an arbitrary normalized many-qubit state $|\Psi\rangle$, which we decompose as:
\begin{equation}
|\Psi\rangle = \Nn_0\big ( |\Psi_0\rangle + |\Psi_0^\bot\rangle\big),
\end{equation}
where $\Nn_0$ is a normalization constant, $\Psi_0$ is the normalized right eigen-vector of $\TM$ corresponding to $\Lambda_0$, denoted before by $\tilde \Psi_0^r$, and $\Psi_0^\bot$ is the orthogonal component of $\Psi$,
\begin{equation}
|\Psi_0^\bot \rangle = \Nn_0^{-1}\big (|\Psi\rangle - \langle \Psi_0|\Psi\rangle \, |\Psi_0 \rangle \big)
\end{equation} 
Note that $\Nn_0$ is given by $\Nn_0 = \langle \Psi_0|\Psi\rangle$ and
\begin{equation}
\|\Psi_0^\bot\| =\big [\tfrac{1}{\langle \Psi_0|\Psi\rangle^2}-1 \big ]^\frac{1}{2}.
\end{equation} 
Next, we observe that, for any $m \geq 0$,
\begin{equation}\label{Eq:R1}
\Cc_{\TM^m} |\Psi \rangle = \Nn_m \big( \Lambda_0^m|\Psi_0\rangle + \TM^m |\Psi_0^\bot\rangle \big),
\end{equation}
where $\Nn_m$ is the normalization constant automatically introduced by the quantum circuit,
\begin{equation}
\begin{aligned}
\Nn_m & = \big [ \Lambda_0^{2m} + \|\TM^m \Psi_0^\bot \|^2 \big ]^{-\frac{1}{2}} \\
 & = \Lambda_0^{-m}\big [ 1 + \|\ \widetilde \TM^m \Psi_0^\bot \|^2 \big ]^{-\frac{1}{2}}.
 \end{aligned}
\end{equation}
Then the action of $\Cc_{\TM^m}$ on the state can be written more conveniently as
\begin{equation}\label{Eq:R2}
\begin{aligned}
 \Cc_{\TM^m} |\Psi \rangle & = \Ff_m \big( | \Psi_0\rangle + \widetilde \TM^m |\Psi_0^\bot\rangle \big), \\
  \Ff_m & = \big [ 1 + \|\ \widetilde \TM^m \Psi_0^\bot \|^2 \big ]^{-\frac{1}{2}}.
\end{aligned}
\end{equation}
Furthermore, the normalization constant can be computed, in theory and practice, as
\begin{equation}
\Ff_m = \langle \Psi_0 | \Cc_{\TM^m}|\Psi \rangle,
\end{equation}
which gives us access to the quantity
\begin{equation}\label{Eq:RM3}
\|\widetilde \TM^m \Psi_0^\bot\| = \big [\Ff_m^{-2} - 1\big ]^\frac{1}{2}
\end{equation}
The interest in this quantity comes from the fact that
\begin{equation}
\|\widetilde \TM^m \Psi_0^\bot\| \leq \lambda_1^m \|\Psi_0^\bot\|, \ \ \forall \ m \geq 1,
\end{equation} 
or
\begin{equation}
\lambda_1 \geq \big [ \|\Psi_0^\bot\|^{-1} \|\widetilde \TM^m \Psi_0^\bot\| \big ]^\frac{1}{m}, \ \ \forall \ m \geq 1,
\end{equation}
with the inequality becoming  equality in the asymptotic limit $m \to \infty$. Collecting the facts, our conclusion is that $\lambda_1$ is always above the graph of the sequence
\begin{equation}
\left [ \frac{\Ff_m^{-2} - 1}{\Ff_0^{-2} - 1} \right ]^\frac{1}{2m}, \quad m \geq 1.
\end{equation}

While we did mentioned that the above sequence actually converges to $\lambda_1$ as $m \to \infty$, one needs to be aware that already from $m=2$, we are dealing with $\Ff_m$'s that are very close to one and, as such, are very difficult to resolve with reasonable numbers of shots. For this reason, our recommendation is to use the following estimator for $\lambda_1$: 
\begin{equation}\label{Eq:Estimator}
{\rm estimator} := \left [ \frac{\Ff_1^{-2} - 1}{\Ff_0^{-2} - 1} \right ]^\frac{1}{2},
\end{equation}
which should give a firm lower bound on $\lambda_1$. This is because $\Cc_T|\Psi\rangle$ can be computed with good precision with a reasonable number of shots and similarly for $|\Psi_0\rangle$, if one uses the second method described in the sub-section~\ref{Sub-sec:Tests} (see the last comment of that sub-section). 

Fig.~\ref{Fig:Lambda1} reports the quantum simulations of the quantity in Eq.~\ref{Eq:Estimator} for the transfer matrices analyzed in Fig.~\ref{Fig:PowersC0p4} and \ref{Fig:PowersC0p0}. In these simulations, the input state was sampled 100 times using a pseudo-random process and $\Psi_0$ was resolved using six iterations of the $\Cc_\TM$ circuit. According to the above discussion and Fig.~\ref{Fig:TEig}, the output of these simulations should be below the known values of $\lambda_1=0.11$ and $0.18$, respectively, and this is certainly the case in Fig.~\ref{Fig:Lambda1} if we take into account the finite precision of the simulations, which involve $10^5$ shots and exclude an insignificant percentage of exceptions. The latter are associated to a poor histograms resulted from low numbers of meaningful shots. 

\section{Conclusions and Outlook}

In this work, we have singled out specific characteristics of the planar vertex models that make them particularly attractive from the quantum computation point of view. To illustrate our observation, we have coded medium size models and demonstrated that we can reproduce the expected results, up to two significant digits, using quantum simulators. In the process, we supplied a generic strategy for implementing non-unitary matrices with quantum circuits containing one acilla qubit and one projective measurement. We also pointed out that the number of qubits and the depth of the quantum circuits grows linearly with the lateral size of the lattice.

From our quantum simulations, we were able to extract the eigen-vector corresponding to the largest eigenvalue of the transfer matrix, a quantity that is paramount for computing the expected values of the physical observables. Working with two different input parameters for the model, we were also able to show that the convergence towards the thermodynamic limit corroborates with the structure of the spectra of the corresponding transfer matrices. In particular, for a variety of input parameters, we found that the thermodynamic limit with respect to the vertical size is achieved after three or four layers. Lastly, we devise a robust procedure to estimate the ratio between the second to first largest eigenvalues, a quantity that determines the asymptotic behavior of the correlation functions.

In the near future, we plan to test the circuits on real quantum computers. For this, we are currently working on optimizing the transpiling. We also plan to code and simulate the expected values of various physical observables of interest and to investigate the correlations by hard coding.

\label{Sec:Appendix1}

\section{Appendix 1}

{\scriptsize
\begin{lstlisting}[language=Python,caption={ \ },label={Script:1}]
##############################################
# implements and tests the action of a       #
# diagonal matrix on a 2-qubit state         #
##############################################
import numpy as np
# generates the D matrix and its unitary lift
d=[]
for i in range(2**2):
    d.append(np.random.rand())
print("D diagonal values:",d)
DM = [[d[0], 0, 0, 0, np.sqrt(1-d[0]*d[0]), 0, 0, 0], 
      [0, d[1], 0, 0, 0, np.sqrt(1-d[1]*d[1]), 0, 0], 
      [0, 0, d[2], 0, 0, 0, np.sqrt(1-d[2]*d[2]), 0],
      [0, 0, 0, d[3], 0, 0, 0, np.sqrt(1-d[3]*d[3])],
      [np.sqrt(1-d[0]*d[0]), 0, 0, 0,-d[0], 0, 0, 0], 
      [0, np.sqrt(1-d[1]*d[1]), 0, 0, 0,-d[1], 0, 0], 
      [0, 0, np.sqrt(1-d[2]*d[2]), 0, 0, 0,-d[2], 0],
      [0, 0, 0, np.sqrt(1-d[3]*d[3]), 0, 0, 0,-d[3]]]
# generates a random normalized 2-qubit state
init_state=[]
for i in range(2**2):
    init_state.append(np.random.rand())
norm=0.0
for w in init_state:
    norm=norm+w*w         
init_state=init_state/np.sqrt(norm)   
print("this is the initial state",init_state)
# builds the quantum circuit
from qiskit import *
q = QuantumRegister(3)
c = ClassicalRegister(3)
qc = QuantumCircuit(q,c)
qc.initialize(init_state,[q[0],q[1]])
qc.unitary(DM,[0,1,2])
qc.barrier()
qc.measure([0,1,2],[0,1,2])
# prepares and executes
from qiskit.compiler import *
from qiskit.providers.aer import *
from qiskit.visualization import *
simulator = QasmSimulator()
compiled_circuit=transpile(qc,simulator)
my_qobj = assemble(compiled_circuit)
runs=10000
job=simulator.run(my_qobj,shots=runs)
result=job.result()
print("length and counts")
counts=result.get_counts()
print(len(counts),counts) 
# collects the relevant histogram
t=list(counts.keys())
u=list(counts.values())
x1=[]
y1=[]
tot=0.0
for v in t:
    if int(v,base=2) <4:
       x1.append(int(v,base=2))
       y1.append(counts[v])
       tot=tot+counts[v]
print("meaningful runs",tot)
print("relevant counts")
print(x1)
print(y1)
for i in range(len(y1)):
    y1[i]=np.sqrt(y1[i]/tot)
print("simulated action of D matrix")
print(x1)
print(y1)
# generates the expected result
x2=[]
y2=[]
norm=0
for i in x1:
    x2.append(i)
    y2.append(d[i]*init_state[i])
    norm=norm+y2[len(y2)-1]**2
y2=y2/np.sqrt(norm)
# compares the histograms (simulated and expected)
print("simulator",y1)
print("expected",y2)
import matplotlib.pyplot as plt
g1=plt.scatter(x1,y1,marker='o',s=50)
g2=plt.scatter(x2,y2,marker='+',s=60)
plt.show()
plt.close()
qc.draw()
\end{lstlisting}

\section{Appendix 2}
\label{Sec:Appendix2}

{\scriptsize
\begin{lstlisting}[language=Python,caption={ \ },label={Script:1}]
#################################
# action of T matrix to a power #
# on a many-qubit state         #
#################################
import numpy as np
beta=2.0
fact=0.4
# generates the R matrix
R=np.zeros((4,4))
fill="0000"
eps=[]
for i in range(16):
    b=np.base_repr(i,base=2)
    q=len(b)
    b=fill+b
    b=b[q:]
    s=0.0
    for j in b:
        s=s+int(j)
    eps.append(fact*s+np.random.rand())
    u1=2*int(b[0])+int(b[1])
    u2=2*int(b[2])+int(b[3])
    R[u1][u2]=np.exp(-beta*eps[i])
# executes the singular value decomposition
U,d,V=np.linalg.svd(R)
d=d/d[0]  # normalizes the SV
print("normalized SV",d)
# generates the 8x8 unitary matrix DM
DM = [[d[0], 0, 0, 0, np.sqrt(1-d[0]*d[0]), 0, 0, 0], 
      [0, d[1], 0, 0, 0, np.sqrt(1-d[1]*d[1]), 0, 0], 
      [0, 0, d[2], 0, 0, 0, np.sqrt(1-d[2]*d[2]), 0],
      [0, 0, 0, d[3], 0, 0, 0, np.sqrt(1-d[3]*d[3])],
      [np.sqrt(1-d[0]*d[0]), 0, 0, 0,-d[0], 0, 0, 0], 
      [0, np.sqrt(1-d[1]*d[1]), 0, 0, 0,-d[1], 0, 0], 
      [0, 0, np.sqrt(1-d[2]*d[2]), 0, 0, 0,-d[2], 0],
      [0, 0, 0, np.sqrt(1-d[3]*d[3]), 0, 0, 0,-d[3]]]
# builds the circuit
from qiskit import *
R_cnt=4
T_power=3
qr=R_cnt+2
cr=R_cnt*T_power+R_cnt+1
init_state=np.zeros(2**qr)
init_state[0]=1.0
print("this is the initial state",init_state)
q = QuantumRegister(qr)
c = ClassicalRegister(cr)
qc = QuantumCircuit(q,c)
qc.initialize(init_state,[q[i] for i in range(qr)])
for i in range(T_power):
    for j in range(R_cnt):
        qc.unitary(V,[R_cnt-j-1,R_cnt],label="V")
        qc.unitary(DM,[R_cnt-j-1,R_cnt,R_cnt+1],label="D")
        qc.unitary(U,[R_cnt-j-1,R_cnt],label="U")
        qc.measure([R_cnt+1],[cr-j-1-i*R_cnt])
        qc.barrier()
qc.measure([j for j in range(R_cnt+1)],[j for j in range(R_cnt+1)])
qc.draw()
# prepares and executes
runs=40000
from qiskit.compiler import *
from qiskit.providers.aer import *
simulator = QasmSimulator()
compiled_circuit=transpile(qc,simulator)
my_qobj = assemble(compiled_circuit)
job=simulator.run(my_qobj,shots=runs)
result=job.result()
counts=result.get_counts()
# processes and outputs the result
t=list(counts.keys())
xs=[]
ys=[]
tot=0.0
for v in t:
    if int(v,base=2) <2**(qr-1):
       xs.append(int(v,base=2))
       ys.append(counts[v])
       tot=tot+counts[v]
print("meaningful runs",tot,"out of total shots",runs)
for i in range(len(ys)):
    ys[i]=np.sqrt(ys[i]/tot)
print("action on the init_state of T to power",T_power)
print(xs)
print(ys)
\end{lstlisting}

\end{document}